\documentclass[11pt]{article}
\usepackage{fullpage}
\usepackage{booktabs} 

\usepackage{amsmath,amsfonts,amssymb,mathtools,amsthm}
\usepackage{natbib}
\usepackage[hidelinks]{hyperref}
\usepackage[T1]{fontenc}
\usepackage{xcolor}
\usepackage{algorithm}
\usepackage{algpseudocode}
\usepackage{graphicx}
\usepackage{enumerate}
\usepackage{caption}
\usepackage{setspace}

\usepackage{tikz}
\usepackage{tkz-euclide}    
\usepackage{pgfplots}       
\usetikzlibrary{shapes,snakes}
\usetikzlibrary{arrows}
\usepackage{subfigure}
\pgfdeclarelayer{background}
\pgfsetlayers{background,main}
\usepackage{subfiles}
\pgfplotsset{compat=1.18}
\usetikzlibrary{trees}
\usetikzlibrary{positioning}
\usepackage{tikz-cd}

\makeatletter
\def\pt@get#1#2{
  \tikz@scan@one@point\pgfutil@firstofone#2\relax%
  \csname pgf@x#1\endcsname=\pgf@x%
  \csname pgf@y#1\endcsname=\pgf@y%
}
\tikzset{
  parabola through/.style={
    to path={{[x={(\pgf@xc,\pgf@yc)}, y=\parabola@y, shift=(\tikztostart)]
      -- (0,0) .. controls (1/3,1/3) and (2/3,1/3) .. (1,0) \tikztonodes}--(\tikztotarget)}
  },
  parabola through/.prefix code={
    \pt@get{a}{(\tikztostart)}\pt@get{b}{#1}\pt@get{c}{(\tikztotarget)}%
    \advance\pgf@xb by-\pgf@xa\advance\pgf@yb by-\pgf@ya%
    \advance\pgf@xc by-\pgf@xa\advance\pgf@yc by-\pgf@ya%
    \pgfmathsetmacro\parabola@y{(\pgf@yc-\pgf@xc/\pgf@xb*\pgf@yb)%
      /(\pgf@xb-\pgf@xc)*\pgf@xc}%
  }
}
\makeatother

\global\long\def\E{\mathbb{E}}

\newtheorem{theorem}{Theorem}
\newtheorem{proposition}[theorem]{Proposition}
\newtheorem{corollary}[theorem]{Corollary}

\newtheorem{assumption}{Assumption}

\theoremstyle{definition}

\newtheorem{definition}[theorem]{Definition}

\usepackage{array}      
\usepackage{multirow}
\newcommand\mc[1]{\multicolumn{1}{c}{#1}} 

\title{Persuasion, Delegation, and Private Information in Algorithm-Assisted Decisions}

\author{Ruqing Xu}
\date{%
Cornell University, Department of Economics\\[2ex]%
February, 2024
}

\begin{document}

\maketitle

\begin{abstract}
    A principal designs an algorithm that generates a publicly observable prediction of a binary state. She must decide whether to act directly based on the prediction or to delegate the decision to an agent with private information but potential misalignment. We study the optimal design of the prediction algorithm and the delegation rule in such environments. Three key findings emerge: (1) Delegation is optimal if and only if the principal would make the same binary decision as the agent had she observed the agent’s information. (2) Providing the most informative algorithm may be suboptimal even if the principal can act on the algorithm’s prediction. Instead, the optimal algorithm may provide more information about one state and restrict information about the other. (3) Well-intentioned policies aiming to provide more information, such as keeping a ``human-in-the-loop'' or requiring maximal prediction accuracy, could strictly worsen decision quality compared to systems with no human or no algorithmic assistance. These findings predict the underperformance of human-machine collaborations if no measures are taken to mitigate common preference misalignment between algorithms and human decision-makers.
\end{abstract}

\section{Introduction}
Decisions today are increasingly complex and data intensive. Toward the goal of making better decisions, decision-aid algorithms have been developed and employed in many consequential domains, including criminal justice, healthcare, and credit lending systems. These algorithms take in data from decision subjects (e.g., defendants, patients, loan applicants) and generate predictions. 
These predictions can either be used to make decisions automatically, or be disclosed to human agents who ultimately make a decision. While the agents may have private knowledge that helps to make a better decision, they may also exhibit behavioral biases, cognitive limitations, or private motives that might misalign with the algorithm. 

A real-life example is the Child Protection Service (CPS) in Allegheny County, PA, which utilizes an algorithm to assign risk scores to reports of child maltreatment \citep{cheng_how_2022}. If the risk score exceeds certain thresholds, cases are automatically screened-in. Otherwise, a social worker sees the score and other available information, then makes the final decision. Suppose that a social worker under-investigates since investigation is costly to him, but he possesses intuition and contextual information that are inaccessible to the algorithm. How should we design the predictive algorithm to collaborate with such a privately informed but biased agent? When should we pass the prediction and the decision to the agent, for the purpose of better information?

Indeed, preference misalignment is widespread in other applications. Bank managers may be more risk-averse than a profit-maximizing algorithm when evaluating high-risk opportunities. Physicians may overweight salient but less predictive signals in diagnosing and testing \citep{mullainathan_diagnosing_2022}. Judges can exhibit behavioral inconsistencies or even biases towards certain demographic groups \citep{eren_emotional_2018, kleinberg_human_2017}. 
If the quality of the final decision is of concern, we ought to take into account the agents' private information and potential misalignment in designing the predictive algorithm, as well as in deciding when the decision authority should be given to an agent.

We present a principal-agent model to study the algorithm design and delegation strategy when the agent is privately informed but potentially misaligned. We model the algorithm as a ``signal'' whose realizations (predictions) convey information about the underlying state. The design of the algorithm corresponds to a persuasion problem where the principal chooses the optimal signal structure. We model the problem of whether to pass the decision to the human as a delegation problem. Our results suggest that delegation 
is strictly valuable if and only if the principal would make the same decision as the agent had she observed the agent’s information. 
We find that providing the most informative algorithm may not be optimal even if the principal can choose to act on the algorithm’s prediction. Moreover, the optimal algorithm may provide more information about one state while restricting information about the other. Lastly, we show that naive policies, such as keeping a ``human-in-the-loop'' or requiring maximal prediction accuracy, strictly worsen decision quality for some decision subjects, even compared to systems with no human or no algorithmic assistance. 
We suggest that the underperformance of human-machine collaborations widely observed in empirical settings \citep{lai2021towards} can be understood through this theoretical lens. 

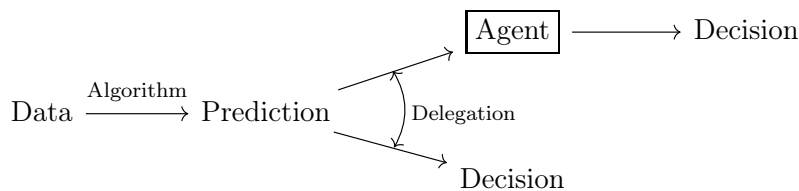
\begin{figure}[htb]
    \centering
    \[
        \begin{tikzcd}[column sep=large, row sep=small]
         && \text{\fbox{Agent}} \arrow[r] &\text{Decision}\\
            \text{Data} \arrow[r, "\text{Algorithm}" {yshift=0.2em}]  & \text{Prediction} \arrow[ur, ""{name=A}] \arrow[dr, ""{name=B, below}]& & \\
            && \text{Decision} &
            \arrow[from=A,to=B, bend left, leftrightarrow, "\text{Delegation}" {yshift=-0.2em}]
        \end{tikzcd}
    \]
    \caption{Diagram of the two decision problems of the principal}
    \label{fig:diagram}
\end{figure}

Formally, we consider a model with binary states and binary actions. The principal and the agent each receive a signal that contains information about the underlying state. The principal's signal is publicly observable while the agent's signal is private to himself.\footnote{\enskip For the ease of modeling, we assume that the principal always reveals her signal to the agent regardless of whether she delegates to the agent, which makes the principal's signal essentially public. It is without loss of generality because the agent cannot use the signal when he is not delegated with the decision.} We allow for arbitrary state-dependent preferences for both players but no contingent transfers. The principal faces two decision problems. First, the principal designs the information structure of the public signal, subject to a constraint on the maximal informativeness. After the principal observes the signal realization, she also decides whether to act with the current information, or to delegate the decision to the agent. 
Although these decisions are made at different times, the principal must take into account the delegation decision in the design of the public signal. Figure~\ref{fig:diagram} illustrates.

We begin by characterizing the necessary and sufficient conditions under which delegation is valuable. Proposition \ref{prop:delegation} proves that delegation strictly improves the principal's payoff if and only if the principal would take the same action as the agent if she were to observe the agent's signal. In all other cases, the principal is either indifferent or strictly prefers to take control due to the agent's misalignment.
We then analyze the comparative statics on the principal's payoffs with respect to the informativeness of the agent's signal and the degree of preference misalignment between the players. Proposition \ref{prop:cs_q} highlights an interesting result: the principal's delegation payoff could strictly decrease when the agent becomes more informative. In particular, this happens when the principal's interim posterior (after observing the public signal) prefers the action that is less preferred by the agent. On the other hand, Proposition \ref{prop:cs_l} shows that the principal's payoff weakly increases when facing a more aligned agent.

After characterizing the optimal delegation strategy, we turn to the principal's problem of designing the optimal public signal. We show in Proposition~\ref{prop:infodesign} that the most informative public signal is optimal if and only if a convexity condition on the posteriors under the maximal signal holds. Otherwise, the optimal signal maximizes information one state while restricting information about the other. Lastly, we highlight in Propositions \ref{prop:welfare1} and \ref{prop:welfare2} that naive policies, such as mandating delegation (keeping a ``human-in-the-loop'') or disregarding persuasion (maximizing prediction accuracy), strictly worsen decision quality for some decision subjects in the absence of perfectly aligned agents and state-revealing signals. 

This paper combines two strands of literature in economics -- optimal delegation and Bayesian persuasion. The classical delegation problem considers a principal who faces a privately informed but misaligned agent. The principal influences the agent's behavior by specifying a set of actions (called the delegation set) from which the agent can choose \citep{holmstrom1978incentives, holmstrom1980theory}. Following that, \citet{alonso_optimal_2008} and \citet{amador2013theory} characterize the optimal delegation set under more general preferences. This paper is also related to the Bayesian persuasion problem \citep{kamenica_bayesian_2011}. The persuasion problem considers the situation where the principal influences the decision of the agent by designing the information structure of a publicly observable signal.

This paper differs from and contributes to the above literature by considering the joint design of optimal persuasion and delegation mechanisms. 
\citet{lou_optimal_2022, vairo2023value} study a similar problem on the joint design of delegation and disclosure rules. However, \citet{lou_optimal_2022} focuses on a setting where the principal designs the structure of the agent's signal, whose realization remains unobservable to the principal. Our paper differs in that the principal designs the public signal instead of the agent's private signal, and can observe the realization of the public signal and choose to use it directly. On the other hand, \citet{vairo2023value} assumes that the agent's private information is about his \emph{type}, which affects the payoff of both players. 
In contrast, we assume that the agent holds private information about the underlying \emph{state}, while the preference of the agent is common knowledge. This difference stems from the context of decision-aid algorithms we hope to model, where both the algorithm and the human agent try to learn about the true state and take the optimal action.  

There have been efforts to model human-machine collaborations from both the fields of economics and computer science. \citet{rastogi_unifying_2022} and \citet{donahue_human-algorithm_2022} investigate the optimal ways to aggregate independent algorithmic and human predictions, and under what conditions these aggregations outperform individual predictions. \citet{straitouri2022improving} proposes an algorithm that selects an optimal set of possible labels and presents it to a human expect for final selection. \citet{xu_decision-aid_2023} and \cite{mclaughlin_algorithmic_2023} study theoretically how to adjust the design of decision-aid algorithms to counteract human biases. \citet{agrawal2018prediction} studies the complementarity of machine predictions and humans' private information about the payoffs of state contingent actions. To our knowledge, this is the first paper to study the design of algorithms and delegation to humans, two elements of human-machine collaboration, in the presence of misaligned agents.

\section{Problem Setting}\label{sec:model}
There are two players: a principal (she) and an agent (he). The state of the world $\Theta$ can take one of the two values, $\theta\in \{0, 1\}$. It can represent ``low'' or ``high,'' ``bad'' or ``good,'' ``guilty'' or ``innocent,'' etc. The principal and the agent share a common prior $\mu$ on the probability of the high state, $\mu = \Pr(\Theta=1)$. Both players receive a publicly observable signal, while the agent also receives a private signal.

The principal chooses and commits to the information structure of the public signal. The information structure of the private signal is exogenous and common knowledge. After observing the public signal, the principal decides whether to take an action with the current information or to delegate the decision to the agent. If delegated, the agent observes both the public and private signals and takes an action.

\subsection{Signals and Informativeness}
We denote the public and private signals as $S_1$ and $S_2$, respectively, with their realizations $s_1$ and $s_2$. The set of possible signals realizations are $\{0, 1\}$. We call $\mu_{s_1} = \Pr(\Theta=1 \mid S_1 = s_1)$ the \emph{interim posterior}, which is the updated belief of the players after observing the public signal. We call $\mu_{s_1s_2} = \Pr(\Theta=1 \mid S_1 = s_1, S_2 = s_2)$ the \emph{final posterior}, which is the agent's posterior after observing both signals.
Signals are generated according to the following information structures:

\medskip 

\begin{minipage}[c]{0.5\textwidth}
  \centering
  $\setlength{\extrarowheight}{2pt}
  \begin{array}{rr|c c}
      &  \mc{}      & \multicolumn{2}{c}{\textup{Signal $S_1$}} \\
      &       & 0 & 1 \\ \cline{2-4}
      \multirow{2}{*}{\textup{State $\Theta$}} & 0             &  p_0 & 1-p_0 \\ 
       & 1            & 1-p_1 &   p_1 \\ 
  \end{array}$
  \captionof{table}{Public signal}
\end{minipage}%
\begin{minipage}[c]{0.5\textwidth}
  \centering
  $\setlength{\extrarowheight}{2pt}
  \begin{array}{rr|c c}
      &  \mc{}      & \multicolumn{2}{c}{\textup{Signal $S_2$}} \\
      &       & 0 & 1 \\ \cline{2-4}
      \multirow{2}{*}{\textup{State $\Theta$}} & 0             &  q_0 & 1-q_0 \\ 
       & 1            & 1-q_1 &  q_1 \\ 
  \end{array}$
  \captionof{table}{Agent's signal}
\end{minipage}
\vspace{1em}

The table entries represent $\Pr(s\mid\theta)$, the conditional probability of observing the signal $s$ given the state $\theta$. We assume that $p_0 + p_1 >1$ and $q_0 + q_1 >1$, meaning that a good (bad) signal is good (bad) news. 

Since we are interested in the principal's payoffs when the signals become more informative, we need a theory to compare the informativeness of information structures. We adopt a standard measure of informativeness in informational economics, Blackwell informativeness \citep{blackwell1951comparison, blackwell1953equivalent}. We say that signal $S$ is (Blackwell) more informative than signal $S'$ if for any prior, the distribution of posterior beliefs after observing $S$ is a mean-preserving spread of the distribution of posterior beliefs after observing $S'$. In the context of binary signals, it is equivalent to saying that the interval between the two posteriors after observing the high and low realizations under $S$ strictly contains the interval between the two posteriors under $S'$. 

Blackwell informativeness is independent of decision-makers' priors and preferences. This means that any Bayesian agent, facing any decision problem, can obtain a higher expected payoff if she receives a signal that is Blackwell more informative. This is desirable in our setting since it allows the comparison of signal informativeness to be independent of the prior or the preferences. However, we also note that the Blackwell order is incomplete -- not all signals can be ranked in terms of Blackwell informativeness.

To make the problem nontrivial, we impose a maximal Blackwell constraint on the information design problem of the principal.\footnote{\enskip If the principal has access to an unrestricted space of signals, she will simply choose the state-revealing signal and never delegate.} The principal can only choose from signals that are equally or less informative than the maximally informative signal (i.e., the maximal signal), denoted by $\lambda$. The posterior beliefs under the maximal signal are denoted by $\mu_0^\lambda$ and $\mu_1^\lambda$.

\subsection{Preferences and Payoffs}
The principal and the agent have access to a common set of actions $Y = \{0, 1\}$. They have (possibly) misaligned preferences that can be represented by two payoff matrices:

\medskip 

\begin{minipage}[c]{0.5\textwidth}
    \centering
    $\setlength{\extrarowheight}{2pt}%
    \begin{array}{rr|c c}
         &  \mc{} & \multicolumn{2}{c}{\textup{Action $Y$}} \\
         &  & 0 & 1 \\ \cline{2-4}
        \multirow{2}{*}{\textup{State $\Theta$}} & 0             &  r_{00} & r_{01} \\ 
         & 1            & r_{10} &  r_{11} \\ 
    \end{array}$
    \captionof{table}{Principal's payoffs}
\end{minipage}%
\begin{minipage}[c]{0.5\textwidth}
    \centering
    $\setlength{\extrarowheight}{2pt}%
    \begin{array}{rr|c c}
         &  \mc{} & \multicolumn{2}{c}{\textup{Action $Y$}} \\
         &  & 0 & 1 \\ \cline{2-4}
        \multirow{2}{*}{\textup{State $\Theta$}} & 0             &  v_{00} & v_{01} \\ 
         & 1            & v_{10} &  v_{11} \\ 
    \end{array}$
    \captionof{table}{Agent's payoffs}
\end{minipage}
\vspace{1em}

The first subscript represents the state whereas the second represents the action. For example, $r_{ij}$ represents the payoff of the principal when under state $i$, action $j$ is taken. On top of the payoff structure, both players are von Neumann-Morgenstern expected utility maximizers. This payoff structure subsumes many commonly assumed payoff functions with binary states and actions, for example, when the principal and the agent have heterogeneous costs of taking certain actions; when the principal and the agent have different benefits of matching the state \citep{kamenica_bayesian_2011}; and when the agent exhibits loss aversion when taking a risky action in a bad state. We impose the following assumption on the payoff structures:
\begin{assumption} 
  The principal's payoffs satisfy $r_{00} > r_{01}$ and $r_{11} > r_{10}$. Similarly, the agent's payoffs satisfy $v_{00} > v_{01}$ and $v_{11} > v_{10}$. This assumption has two implications:
\begin{enumerate}[(1)]
  \item There is no dominant action (i.e., action one would take regardless of the state) for either player.  
  \item Each player's payoff function of action $1$ crosses their payoff function of action $0$ from below. This means that both players switch from action $0$ to action $1$ at some point as their beliefs about state $1$ increase. 
\end{enumerate}\label{ass:payoff}
\end{assumption}

We show in Appendix~\ref{app:assumption1} that this assumption makes the problem interesting, since the principal never strictly prefers to delegate to the agent when this assumption is violated and the information design problem is trivially solved by choosing the maximally informative signal.  

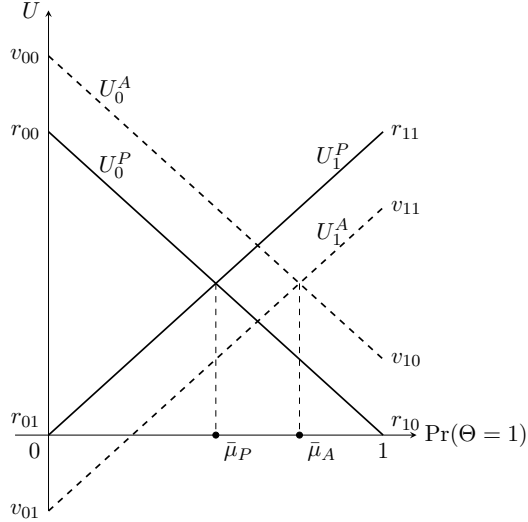
\begin{figure}[h]
  \centering 
  \scalebox{0.8}{%
  \begin{tikzpicture}
    \pgfplotsset{compat=1.17}
    \begin{axis}[
      xlabel={$\Pr(\Theta=1)$},
      ylabel={$U$},
      xmin=-0.1, xmax=1.1,
      ymax=1.4,
      axis lines=middle,
      samples=100,
      width=0.5\textwidth,
      height=0.6\textwidth,
      xtick=\empty,
      ytick=\empty,
      xlabel style={anchor=west},
      ylabel style={anchor=east},
      clip=false,
      ]
      \addplot[black, thick, domain=0:1] {x};
      \addplot[black, thick, domain=0:1] {1 - x};
      \addplot[black, dashed, thick, domain=0:1] {x - 0.25};
      \addplot[black, dashed, thick, domain=0:1] {1.25-x};
      
      \coordinate (A) at (0.5, 0.5);
      \coordinate (B) at (0.75, 0.5);
      \coordinate (C) at (0.5, 0);
      \coordinate (D) at (0.75, 0);
      
      \node[draw,circle,inner sep=1pt,fill=black] at (C) {};
      \node[draw,circle,inner sep=1pt,fill=black] at (D) {};

      \node[below left] at (0, 0) {$0$};
      \node[below] at (1, 0) {1};
      \draw[dashed] (A) -- (0.5, 0);
      \node[below right] at (0.5, 0) {$\bar{\mu}_P$};
      \draw[dashed] (B) -- (0.75, 0);
      \node[below right] at (0.75, 0) {$\bar{\mu}_A$};

      \node[above] at (0.85, 0.85) {$U^P_{1}$};
      \node[above] at (0.85, 0.6) {$U^A_{1}$};
      \node[above] at (0.2,0.82) {$U^P_{0}$};
      \node[above] at (0.2,1.07) {$U^A_{0}$};

      \node[left] at (0,1) {$r_{00}$};
      \node[left] at (0,1.25) {$v_{00}$};
      \node[left] at (0,0.05) {$r_{01}$};
      \node[left] at (0,-0.25) {$v_{01}$};
      \node[right] at (1,1) {$r_{11}$};
      \node[right] at (1,0.75) {$v_{11}$};
      \node[right] at (1,0.04) {$r_{10}$};
      \node[right] at (1,0.25) {$v_{10}$};

    \end{axis}
  \end{tikzpicture}
  }
  \caption{Players' payoffs as functions of the belief}
  \label{fig:basic}
\end{figure}

In Figure~\ref{fig:basic}, we plot an example of the two players' payoff functions with respect to the players' beliefs about the underlying state. 

For example, $U^P_0(\cdot)$ denotes the \textbf{P}rincipal's payoff when taking action \textbf{0}, as a function of the belief. The four payoff functions are linear combinations of the discrete payoffs.
Assumption \ref{ass:payoff} guarantees that the payoff functions have intersections. The intersection points $\bar{\mu}_P$ and $\bar{\mu}_A$ represent the cutoff points at which the principal and the agent switch to the higher action, respectively. We call the interval between $\bar{\mu}_P$ and $\bar{\mu}_A$ the \emph{disagreement interval (I)}, since the players prefer different actions in this interval. Note that in this example, $\bar{\mu}_P < \bar{\mu}_A$, meaning that the principal prefers to take action $1$ at a wider range of beliefs compared to the agent. Therefore, we say that action $1$ is the \emph{principal-preferred} action.

\begin{figure}[htb]
  \centering
  \subfigure[Principal's non-delegation envelope ($V_N$)\label{fig:envelope_N}]{  \begin{tikzpicture}[scale=0.8, transform shape]
    \begin{axis}[
      xlabel={$\Pr(\Theta=1)$},
      ylabel={$U$},
      xmin=-0.1, xmax=1,
      axis lines=middle,
      samples=100,
      xtick=\empty,
      ytick=\empty,
      xlabel style={anchor=west},
      ylabel style={anchor=east},
      clip=false,
      ]
      \addplot[black, thick, domain=0.5:1] {x};
      \addplot[black, dashed, domain=0:0.5] {x};
      \addplot[black, thick, domain=0:0.5] {1 - x};
      \addplot[black, dashed, domain=0.5:1] {1 - x};

      \coordinate (A) at (0.5, 0.5);
      \coordinate (B) at (0.75, 0.75);
      \coordinate (C) at (0.5, 0);
      \coordinate (D) at (0.75, 0);
      
      \node[draw,circle,inner sep=1pt,fill=black] at (C) {};

      \node[below left] at (0, 0) {$0$};
      \node[below] at (1, 0) {1};
      \draw[dashed] (A) -- (0.5, 0);
      \node[below right] at (0.5, 0) {$\bar{\mu}_P$};

    \end{axis}
  \end{tikzpicture}}
  \subfigure[Principal's delegation envelope ($V_D$)\label{fig:envelope_D}]{\begin{tikzpicture}[scale=0.8, transform shape]
  \begin{axis}[
    xlabel={$\Pr(\Theta=1)$},
    ylabel={$U$},
    xmin=-0.1, xmax=1,
    axis lines=middle,
    samples=100,
    xtick=\empty,
    ytick=\empty,
    xlabel style={anchor=west},
    ylabel style={anchor=east},
    clip=false,
    ]
    \addplot[black, thick, domain=0.75:1] {x};
    \addplot[black, dashed, domain=0:0.75] {x};
    \addplot[black, thick, domain=0:0.75] {1 - x};
    \addplot[black, dashed, domain=0.75:1] {1 - x};

    \coordinate (A) at (0.5, 0.5);
    \coordinate (B) at (0.75, 0.75);
    \coordinate (C) at (0.5, 0);
    \coordinate (D) at (0.75, 0);
    
    \node[draw,circle,inner sep=1pt,fill=black] at (C) {};
    \node[draw,circle,inner sep=1pt,fill=black] at (D) {};

    \node[below left] at (0, 0) {$0$};
    \node[below] at (1, 0) {1};
    \draw[dashed] (A) -- (0.5, 0);
    \node[below right] at (0.5, 0) {$\bar{\mu}_P$};
    \draw[dashed] (B) -- (0.75, 0);
    \node[below right] at (0.75, 0) {$\bar{\mu}_A$};
    \node[above] at (0.63, 0) {$I$};

  \end{axis}
\end{tikzpicture}}
  \caption{Comparison of the principal's payoff envelopes}\label{fig:envelope}
\end{figure}
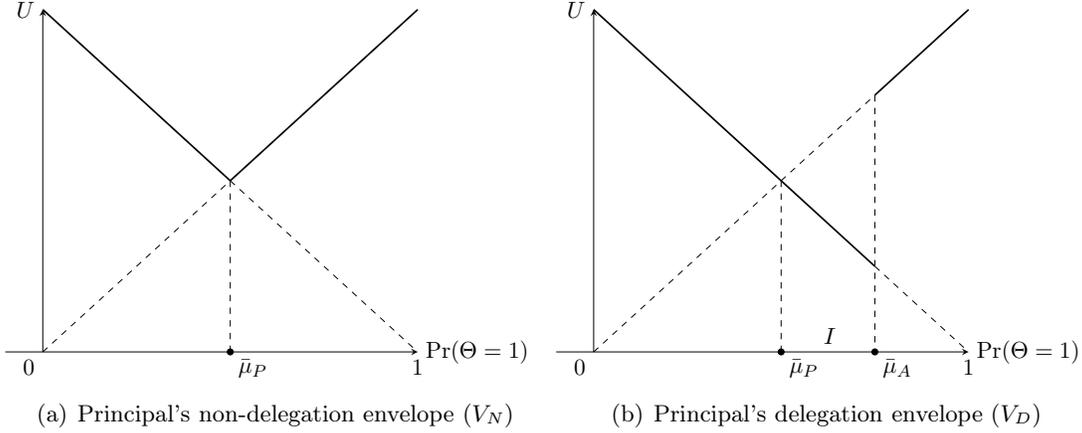

Our analysis greatly simplifies if we look at the \emph{payoff envelopes} of the principal. If the principal does not delegate, the envelope is simply the maximum of $ U^P_0$ and $ U^P_1$, which is continuous and has a kink at $\bar{\mu}_P$. If the principal delegates, however, the agent makes the final decision. Thus, the principal's payoff follows the cutoff point of the agent and becomes discontinuous at $\bar{\mu}_A$.\footnote{\enskip For simplicity of stating theorems, we assume that the agent takes the principal-preferred action when indifferent, so the delegation envelope evaluates to the top point at the discontinuity.}
We denote them as the \emph{non-delegation envelope} ($V_N$) and \emph{delegation envelope} ($V_D$), respectively, as shown in Figure \ref{fig:envelope}. 

\subsection{Design Choices}

\paragraph{Conditioning delegation on the public signal.} In this model, we assume that the principal can condition the delegation decision on the realization of the public signal. Alternatively, one may assume that the principal must commit to delegation or direct action before observing the public signal. We note that these two models correspond to different decision problems: the former captures the decision-making process for each decision subject, whereas the latter chooses between a fully automated system or a human-in-the-loop system for all decisions.

\paragraph{Heterogeneity of preferences and signals for different decision subjects.} One may be concerned with how generalizable our results are when the players' preferences and signal informativeness may change with respect to different decision subjects. However, we believe that this does not limit the overall applicability of our findings. This is because the decision-making process for each decision subject can be repeated under this framework, but with varied parameters for preferences, signal structures of the agent, and maximal Blackwell constraints.

\paragraph{Complete information about the agent's preferences and informativeness.} Common to theoretical work in persuasion and delegation, this paper makes the assumption that the preferences of both players and the information structure of the signals are common knowledge. Such assumptions allow us to cleanly characterize the optimal persuasion and delegation mechanisms.
We recognize that a promising extension to this paper is to develop methods for separately identifying and estimating human agents' biases and information. We discuss this in more detail in the future work section.

\section{Value of Delegation}
In this section, we study the principal's optimal delegation decision after observing the realization of the public signal. Proposition \ref{prop:delegation} characterizes the necessary and sufficient condition for delegation to be strictly valuable.

\begin{proposition}[Necessary and sufficient condition for strict delegation]\label{prop:delegation}
  Given a public signal realization $s_1$ and the associated interim posterior $\mu_{s_1}$, delegation is strictly valuable to the principal if and only if the agent's final posteriors $\mu_{s_{1}0}$ and $\mu_{s_{1}1}$ lie on the opposing sides of the disagreement interval. In other words,
  \[ \mu_{s_{1}0} < \{\bar{\mu}_P, \bar{\mu}_A\} < \mu_{s_{1}1}. \]
\end{proposition}

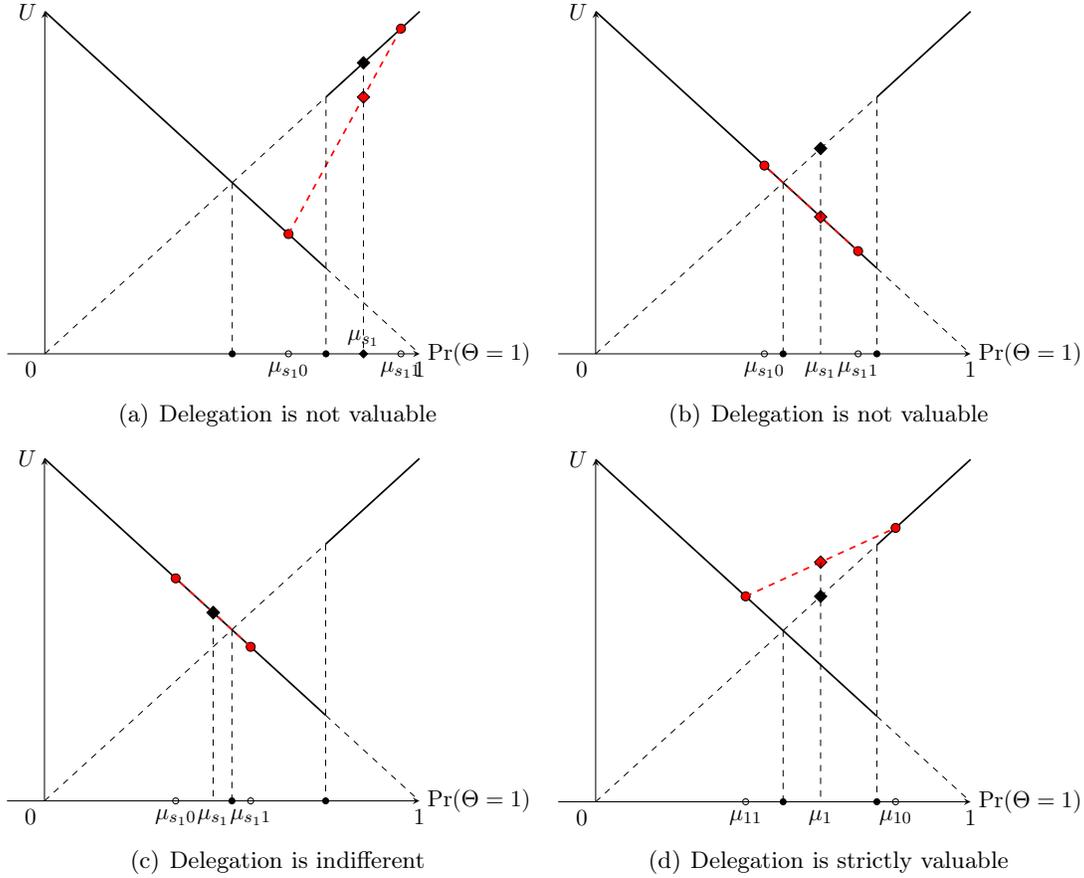
\begin{figure}[htb]
  \centering
  \subfigure[Delegation is not valuable\label{fig:valuable1}]{  \begin{tikzpicture}[scale=0.8, transform shape]
    \pgfplotsset{compat=1.17}
    \begin{axis}[
      xlabel={$\Pr(\Theta=1)$},
      ylabel={$U$},
      xmin=-0.1, xmax=1,
      axis lines=middle,
      samples=100,
      xtick=\empty,
      ytick=\empty,
      xlabel style={anchor=west},
      ylabel style={anchor=east},
      clip=false,
      ]
      \addplot[black, thick, domain=0.75:1] {x};
      \addplot[black, dashed, domain=0:0.75] {x};
      \addplot[black, thick, domain=0:0.75] {1 - x};
      \addplot[black, dashed, domain=0.75:1] {1 - x};

      \coordinate (A) at (0.5, 0.5);
      \coordinate (B) at (0.75, 0.75);
      \coordinate (C) at (0.5, 0);
      \coordinate (D) at (0.75, 0);
      \coordinate (D1) at (0.75, 0);
      \coordinate (E1) at (0.65, 0);
      \coordinate (F1) at (0.95, 0);
      \coordinate (E2) at (0.65,0.35);
      \coordinate (F2) at (0.95,0.95);
      \coordinate (MEAN2) at (0.85,0.85);
      \coordinate (MEAN1) at (0.85,0.75);

      \node[draw,circle,inner sep=1pt,fill=black] at (C) {};
      \node[draw,circle,inner sep=1pt,fill=black] at (D) {};
      \node[draw,circle,inner sep=1pt,fill=none] at (E1) {};
      \node[draw,circle,inner sep=1pt,fill=none] at (F1) {};
      \node[draw,circle,inner sep=1.5pt,fill=red] at (E2) {};
      \node[draw,circle,inner sep=1.5pt,fill=red] at (F2) {};
      \draw[thick, red, dashed] (E2) -- (F2);
      \node[draw,diamond,inner sep=1.5pt,fill=red] at (MEAN1) {};
      \node[draw,diamond,inner sep=1.5pt,fill=black] at (MEAN2) {};
      \node[draw,diamond,inner sep=1pt,fill=black] at (0.85,0) {};
      
      \node[below left] at (0, 0) {$0$};
      \node[below] at (1, 0) {1};
      \draw[dashed] (A) -- (0.5, 0);
      \draw[dashed] (B) -- (0.75, 0);
      \node[below] at (E1) {$\mu_{s_{1}0}$};
      \node[below] at (F1) {$\mu_{s_{1}1}$};
      \draw[dashed] (MEAN2) -- (0.85,0);
      \node[above] at (0.85,0) {$\mu_{s_{1}}$};

    \end{axis}
  \end{tikzpicture}}
  \subfigure[Delegation is not valuable\label{fig:valuable2}]{  \begin{tikzpicture}[scale=0.8, transform shape]
    \pgfplotsset{compat=1.17}
    \begin{axis}[
      xlabel={$\Pr(\Theta=1)$},
      ylabel={$U$},
      xmin=-0.1, xmax=1,
      axis lines=middle,
      samples=100,
      xtick=\empty,
      ytick=\empty,
      xlabel style={anchor=west},
      ylabel style={anchor=east},
      clip=false,
      ]
      \addplot[black, thick, domain=0.75:1] {x};
      \addplot[black, dashed, domain=0:0.75] {x};
      \addplot[black, thick, domain=0:0.75] {1 - x};
      \addplot[black, dashed, domain=0.75:1] {1 - x};

      \coordinate (A) at (0.5, 0.5);
      \coordinate (B) at (0.75, 0.75);
      \coordinate (C) at (0.5, 0);
      \coordinate (D) at (0.75, 0);
      \coordinate (D1) at (0.75, 0);
      \coordinate (E1) at (0.7, 0);
      \coordinate (F1) at (0.45, 0);
      \coordinate (E2) at (0.7,0.3);
      \coordinate (F2) at (0.45,0.55);
      \coordinate (MEAN1) at (0.6,0.4);
      \coordinate (MEAN2) at (0.6,0.6);

      \node[draw,circle,inner sep=1pt,fill=black] at (C) {};
      \node[draw,circle,inner sep=1pt,fill=black] at (D) {};
      \node[draw,circle,inner sep=1pt,fill=none] at (E1) {};
      \node[draw,circle,inner sep=1pt,fill=none] at (F1) {};
      \node[draw,circle,inner sep=1.5pt,fill=red] at (E2) {};
      \node[draw,circle,inner sep=1.5pt,fill=red] at (F2) {};
      \draw[thick, red, dashed] (E2) -- (F2);
      \node[draw,diamond,inner sep=1.5pt,fill=red] at (MEAN1) {};
      \node[draw,diamond,inner sep=1.5pt,fill=black] at (MEAN2) {};

      \node[below left] at (0, 0) {$0$};
      \node[below] at (1, 0) {1};
      \draw[dashed] (A) -- (0.5, 0);
      \draw[dashed] (B) -- (0.75, 0);
      \node[below] at (E1) {$\mu_{s_{1}1}$};
      \node[below] at (F1) {$\mu_{s_{1}0}$};
      \draw[dashed] (MEAN2) -- (0.6, 0);
      \node[below] at (0.6, 0) {$\mu_{s_{1}}$};

    \end{axis}
  \end{tikzpicture}}
  \subfigure[Delegation is indifferent\label{fig:valuable3}]{  \begin{tikzpicture}[scale=0.8, transform shape]
    \pgfplotsset{compat=1.17}
    \begin{axis}[
      xlabel={$\Pr(\Theta=1)$},
      ylabel={$U$},
      xmin=-0.1, xmax=1,
      axis lines=middle,
      samples=100,
      xtick=\empty,
      ytick=\empty,
      xlabel style={anchor=west},
      ylabel style={anchor=east},
      clip=false,
      ]
      \addplot[black, thick, domain=0.75:1] {x};
      \addplot[black, dashed, domain=0:0.75] {x};
      \addplot[black, thick, domain=0:0.75] {1 - x};
      \addplot[black, dashed, domain=0.75:1] {1 - x};

      \coordinate (A) at (0.5, 0.5);
      \coordinate (B) at (0.75, 0.75);
      \coordinate (C) at (0.5, 0);
      \coordinate (D) at (0.75, 0);
      \coordinate (E1) at (0.55, 0);
      \coordinate (F1) at (0.35, 0);
      \coordinate (E2) at (0.55,0.45);
      \coordinate (F2) at (0.35,0.65);
      \coordinate (MEAN1) at (0.45,0.55);
      \coordinate (MEAN2) at (0.45,0.55);

      \node[draw,circle,inner sep=1pt,fill=black] at (C) {};
      \node[draw,circle,inner sep=1pt,fill=black] at (D) {};
      \node[draw,circle,inner sep=1pt,fill=none] at (E1) {};
      \node[draw,circle,inner sep=1pt,fill=none] at (F1) {};
      \node[draw,circle,inner sep=1.5pt,fill=red] at (E2) {};
      \node[draw,circle,inner sep=1.5pt,fill=red] at (F2) {};
      \draw[thick, red, dashed] (E2) -- (F2);
      \node[draw,diamond,inner sep=1.5pt,fill=red] at (MEAN1) {};
      \node[draw,diamond,inner sep=1.5pt,fill=black] at (MEAN2) {};

      \node[below left] at (0, 0) {$0$};
      \node[below] at (1, 0) {1};
      \draw[dashed] (A) -- (0.5, 0);
      \draw[dashed] (B) -- (0.75, 0);
      \node[below] at (E1) {$\mu_{s_{1}1}$};
      \node[below] at (F1) {$\mu_{s_{1}0}$};
      \draw[dashed] (MEAN2) -- (0.45, 0);
      \node[below] at (0.45, 0) {$\mu_{s_{1}}$};

    \end{axis}
  \end{tikzpicture}}
  \subfigure[Delegation is strictly valuable\label{fig:valuable4}]{  \begin{tikzpicture}[scale=0.8, transform shape]
    \pgfplotsset{compat=1.17}
    \begin{axis}[
      xlabel={$\Pr(\Theta=1)$},
      ylabel={$U$},
      xmin=-0.1, xmax=1,
      axis lines=middle,
      samples=100,
      xtick=\empty,
      ytick=\empty,
      xlabel style={anchor=west},
      ylabel style={anchor=east},
      clip=false,
      ]
      \addplot[black, thick, domain=0.75:1] {x};
      \addplot[black, dashed, domain=0:0.75] {x};
      \addplot[black, thick, domain=0:0.75] {1 - x};
      \addplot[black, dashed, domain=0.75:1] {1 - x};

      \coordinate (A) at (0.5, 0.5);
      \coordinate (B) at (0.75, 0.75);
      \coordinate (C) at (0.5, 0);
      \coordinate (D) at (0.75, 0);
      \coordinate (E1) at (0.4, 0);
      \coordinate (F1) at (0.8, 0);
      \coordinate (E2) at (0.4,0.6);
      \coordinate (F2) at (0.8,0.8);
      \coordinate (MEAN1) at (0.6,0.7);
      \coordinate (MEAN2) at (0.6,0.6);

      \node[draw,circle,inner sep=1pt,fill=black] at (C) {};
      \node[draw,circle,inner sep=1pt,fill=black] at (D) {};
      \node[draw,circle,inner sep=1pt,fill=none] at (E1) {};
      \node[draw,circle,inner sep=1pt,fill=none] at (F1) {};
      \node[draw,circle,inner sep=1.5pt,fill=red] at (E2) {};
      \node[draw,circle,inner sep=1.5pt,fill=red] at (F2) {};
      \draw[thick, red, dashed] (E2) -- (F2);
      \node[draw,diamond,inner sep=1.5pt,fill=red] at (MEAN1) {};
      \node[draw,diamond,inner sep=1.5pt,fill=black] at (MEAN2) {};

      \node[below left] at (0, 0) {$0$};
      \node[below] at (1, 0) {1};
      \draw[dashed] (A) -- (0.5, 0);
      \draw[dashed] (B) -- (0.75, 0);
      \node[below] at (E1) {$\mu_{11}$};
      \node[below] at (F1) {$\mu_{10}$};
      \draw[dashed] (MEAN1) -- (0.6, 0);
      \node[below] at (0.6, 0) {$\mu_1$};

    \end{axis}
  \end{tikzpicture}}
  \caption{Value of delegation}\label{fig:valuable}
\end{figure}

\begin{proof}
  We first prove the ``if'' direction. By Bayes' rule, the expected posteriors equals the prior, i.e., $\E_{s_2}[\mu_{s_1s_2}] = \mu_{s_1}$. Since the expected payoff of delegation, $\E_{s_2}[V_D(\mu_{s_1s_2})]$, is linear in beliefs, it equals the linear combination of $V_D(\mu_{s_10})$ and $V_D(\mu_{s_11})$ evaluated at the prior $\mu_{s_1}$. Therefore, if the agent's posteriors lie on the opposing side of the disagreement interval, any linear combination of them is above the principal's non-delegation envelope $V_N$. Thus, at any prior $\mu_{s_1}$, delegation would give strictly higher payoff than non-delegation, i.e., delegation is strictly valuable. 

  We now prove the ``only if'' direction. For the sake of contradiction, suppose that $\mu_{s_{1}0}$ and $\mu_{s_{1}1}$ are not on the opposing sides of the disagreement interval. First observe that delegation cannot be strictly valuable whenever the points $(\mu_{s_10},V_D(\mu_{s_10}))$, $(\mu_{s_11},V_D(\mu_{s_11}))$, and $(\mu_{s_1},V_D(\mu_{s_1}))$ are co-linear. Since if that is the case,
  \[\E_{s_2}[V_D(\mu_{s_1s_2})] = V_D(\E_{s_2}[\mu_{s_1s_2}]) \leq  V_N(\E_{s_2}[\mu_{s_1s_2}]).\]
  The inequality results from the fact that the delegation envelope lies everywhere weakly below the non-delegation envelope. The only other case where the posteriors are not on the opposing side of the disagreement interval and are not co-linear with the prior is as shown in Figure \ref{fig:valuable1}. The expected delegation payoff is a linear combination with one posterior in the disagreement interval and another posterior across the discontinuity at $\bar{\mu}_A$. In this case, delegation makes the principal strictly worse off. 
\end{proof}

Intuitively, Proposition \ref{prop:delegation} shows that the principal strictly prefers to delegate if and only if at the principal's interim posterior, the agent's signal can potentially induce two distinct actions that the principal would agree with if she were to know the agent's signal. To achieve this, the principal's interim posterior cannot be too confident such that the agent's signal cannot provide any ``surprise.'' The agent's signal also needs to be informative enough compared to his misaligned incentives, which is highlighted in the following Corollary.

\begin{corollary}[Necessary condition for strict delegation]
  Given a public signal realization $s_1$ and the associated interim posterior $\mu_{s_1}$, for delegation to be strictly valuable the following must be true: 
  \[\mu_{s_{1}1} - \mu_{s_{1}0} > |\bar{\mu}_P - \bar{\mu}_A| = |I|, \]
  that is, the distance between the agent's final posteriors must be more than the length of the disagreement interval. 
\end{corollary}

\begin{proof}
  It follows immediately from Proposition \ref{prop:delegation}.
\end{proof}

This Corollary captures the intuitive trade-off between the agent's private information and preference misalignment. A necessary condition for the principal to delegate is that the agent's signal is informative enough (the distance between posteriors) compared to the degree of preference misalignment (the length of the disagreement interval). 

\section{Comparative Statics on Delegation}
This section presents the comparative statics results on the principal's payoffs with respect to a change in agent's informativeness or misalignment.

\subsection{Informativeness of the Agent's Signal}
When the principal decides whether to delegate, one important factor is the informativeness of the agent's signal. One may think that the principal is more likely to delegate to an agent with a bigger informational advantage. Surprising, this intuition does not always hold.

As a signal becomes more informative, Blackwell's theorem predicts that, for any prior, at least one of the two posteriors under the new signal will move further away from the prior compared to the old signal. We are interested in the comparative statics effect of increasing the agent's informativeness on two quantities. The first is the principal's \emph{delegation payoff}, i.e., the payoff if she delegates to the agent at a given interim posterior. The second is the principal's \emph{payoff at the optimal delegation decision}, i.e., the payoff when the she chooses optimally between delegation and taking the utility-maximizing action. 

We show that the impact of a more informed agent depends on the location of the interim posterior. In some regions, the principal's delegation payoff can strictly decrease when the agent becomes more informed. 

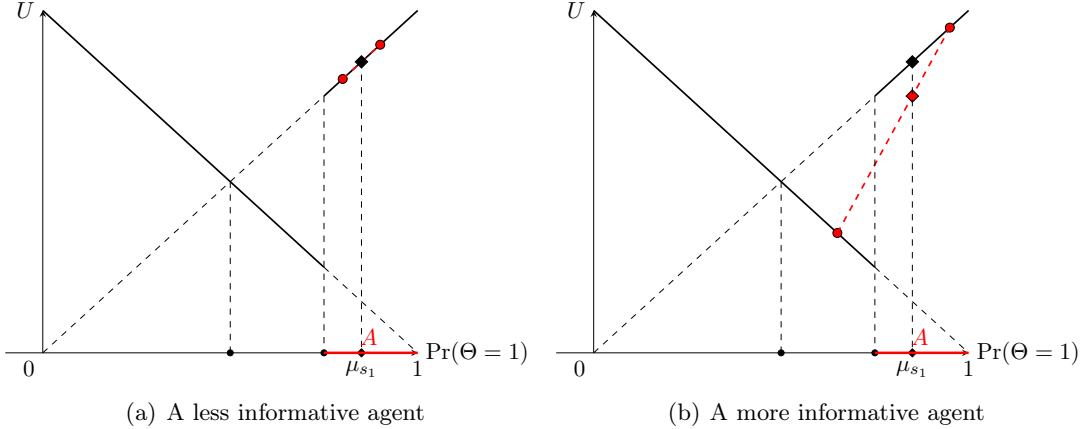
\begin{figure}[htb]
  \centering
  \subfigure[A less informative agent\label{fig:CS1}]{  \begin{tikzpicture}[scale=0.8, transform shape]
    \pgfplotsset{compat=1.17}
    \begin{axis}[
      xlabel={$\Pr(\Theta=1)$},
      ylabel={$U$},
      xmin=-0.1, xmax=1,
      axis lines=middle,
      samples=100,
      xtick=\empty,
      ytick=\empty,
      xlabel style={anchor=west},
      ylabel style={anchor=east},
      clip=false,
      ]
      \addplot[black, thick, domain=0.75:1] {x};
      \addplot[black, dashed, domain=0:0.75] {x};
      \addplot[black, thick, domain=0:0.75] {1 - x};
      \addplot[black, dashed, domain=0.75:1] {1 - x};

      \coordinate (A) at (0.5, 0.5);
      \coordinate (B) at (0.75, 0.75);
      \coordinate (C) at (0.5, 0);
      \coordinate (D) at (0.75, 0);
      \coordinate (D1) at (0.75, 0);
      \coordinate (E1) at (0.8, 0);
      \coordinate (F1) at (0.9, 0);
      \coordinate (E2) at (0.8,0.8);
      \coordinate (F2) at (0.9,0.9);
      \coordinate (MEAN2) at (0.85,0.85);
      \coordinate (MEAN1) at (0.85,0.75);

      \node[draw,circle,inner sep=1pt,fill=black] at (C) {};
      \node[draw,circle,inner sep=1pt,fill=black] at (D) {};
      \node[draw,circle,inner sep=1.5pt,fill=red] at (E2) {};
      \node[draw,circle,inner sep=1.5pt,fill=red] at (F2) {};
      \draw[thick, red, dashed] (E2) -- (F2);
      \node[draw,diamond,inner sep=1.5pt,fill=black] at (MEAN2) {};
      \node[draw,diamond,inner sep=1pt,fill=black] at (0.85,0) {};
      
      \node[below left] at (0, 0) {$0$};
      \node[below] at (1, 0) {1};
      \draw[dashed] (A) -- (0.5, 0);
      \draw[dashed] (B) -- (0.75, 0);
      \draw[dashed] (MEAN2) -- (0.85,0);
      \node[below] at (0.85,0) {$\mu_{s_{1}}$};
    \draw[very thick, red] (0.75,0) -- (1, 0);
        \node[above] at (0.87, 0) {\textcolor{red}{$A$}};
    \end{axis}
  \end{tikzpicture}}
  \subfigure[A more informative agent\label{fig:CS2}]{  \begin{tikzpicture}[scale=0.8, transform shape]
    \pgfplotsset{compat=1.17}
    \begin{axis}[
      xlabel={$\Pr(\Theta=1)$},
      ylabel={$U$},
      xmin=-0.1, xmax=1,
      axis lines=middle,
      samples=100,
      xtick=\empty,
      ytick=\empty,
      xlabel style={anchor=west},
      ylabel style={anchor=east},
      clip=false,
      ]
      \addplot[black, thick, domain=0.75:1] {x};
      \addplot[black, dashed, domain=0:0.75] {x};
      \addplot[black, thick, domain=0:0.75] {1 - x};
      \addplot[black, dashed, domain=0.75:1] {1 - x};

      \coordinate (A) at (0.5, 0.5);
      \coordinate (B) at (0.75, 0.75);
      \coordinate (C) at (0.5, 0);
      \coordinate (D) at (0.75, 0);
      \coordinate (D1) at (0.75, 0);
      \coordinate (E1) at (0.65, 0);
      \coordinate (F1) at (0.95, 0);
      \coordinate (E2) at (0.65,0.35);
      \coordinate (F2) at (0.95,0.95);
      \coordinate (MEAN2) at (0.85,0.85);
      \coordinate (MEAN1) at (0.85,0.75);

      \node[draw,circle,inner sep=1pt,fill=black] at (C) {};
      \node[draw,circle,inner sep=1pt,fill=black] at (D) {};
      \node[draw,circle,inner sep=1.5pt,fill=red] at (E2) {};
      \node[draw,circle,inner sep=1.5pt,fill=red] at (F2) {};
      \draw[thick, red, dashed] (E2) -- (F2);
      \node[draw,diamond,inner sep=1.5pt,fill=red] at (MEAN1) {};
      \node[draw,diamond,inner sep=1.5pt,fill=black] at (MEAN2) {};
      \node[draw,diamond,inner sep=1pt,fill=black] at (0.85,0) {};
      
      \node[below left] at (0, 0) {$0$};
      \node[below] at (1, 0) {1};
      \draw[dashed] (A) -- (0.5, 0);
      \draw[dashed] (B) -- (0.75, 0);
      \draw[dashed] (MEAN2) -- (0.85,0);
      \node[below] at (0.85,0) {$\mu_{s_{1}}$};
    \draw[very thick, red] (0.75,0) -- (1, 0);
        \node[above] at (0.87, 0) {\textcolor{red}{$A$}};
    \end{axis}
  \end{tikzpicture}}
  \caption{Delegation payoff (red diamond point) when agent becomes more informative}\label{fig:CS_agent}
\end{figure}

\begin{proposition}\label{prop:cs_q}
  When the principal's interim posterior is such that both players agree to take the principal-preferred action, changing the agent's signal from not very informative to moderately informative can strictly reduce the principal's payoff of delegation. 
\end{proposition}

\begin{proof}
  We show that there are a class of situations in which Proposition~\ref{prop:cs_q} would apply. Denote as $A$ the range of interim posteriors in which both players agree to take the principal-preferred action. Suppose that $\mu_{s_1} \in A$, and the agent currently receives a signal $S_2$, which induces final posteriors $\mu_{s_10}$ and $\mu_{s_11}$. Consider the case where $\mu_{s_10}, \mu_{s_11} \in A$, i.e., when $S_2$ is not too informative so that the agent's final posteriors are not far from the interim posterior. In this case, the final posteriors are co-linear with the interim posterior, so the principal receives the same payoff for delegation and taking the optimal action. 
  
  Suppose now the agent receives a signal $S'_2$ that is Blackwell more informative. By Blackwell's theorem, at least one of the final posteriors $\mu'_{s_10}$ and $\mu'_{s_11}$ would be strictly further away from the interim posterior $\mu_{s_1}$. Consider $S'_2$ that moves out the posterior closer to the disagreement interval $(I)$ (such a signal must exist once the other posterior arrive at the end point and thus cannot be moved anymore). When the two final posteriors are still co-linear with the interim posterior, the payoff of delegation does not change for the principal. However, once one of the posterior falls into the disagreement interval, the principal's expected payoff of delegation falls below the non-delegation envelope (as shown in Figure~\ref{fig:CS1}-\ref{fig:CS2}). 
  
We show that there exist cases in which delegating to a more informed agent is strictly worse off for the principal. However, this case only happens when the interim posterior lies in the interval $A$ and when the agent's signal changes from not very informative to moderately informative. Even when $\mu_{s_1} \in A$, once $S_2$ becomes informative enough so that neither of the posteriors is in the disagreement interval, an further increase in $S_2$'s informativeness would strictly increase the payoff of delegation. In addition, when $\mu_{s_1} \notin A$, increasing the informativeness of $S_2$ always weakly increases the delegation payoff. 
\end{proof} 

This proposition presents an interesting result: holding everything else equal, the principal's delegation payoff may decrease when facing a more informed agent. In particular, this happens when the principal's interim posterior are such that both players agree to take the principal-preferred action (i.e., the action less preferred by the agent). In this case, the risk of the agent switching to the suboptimal action for the principal upon receiving a somewhat informative signal outweighs the benefit of the signal. It is worth noting that this result echoes \citet{alonso_optimal_2008}, in which they show that the principal may give less discretion to an agent with a bigger informational advantage in some situations.

In contrast, we note that the principal's payoff at the optimal delegation decision is unaffected. This is precisely because the principal is indifferent between delegating or not before the agent's signal increases in informativeness, so she can always fall back to not delegating and ensure the same payoff as before. 

\begin{proposition}\label{prop:cs_q2}
  At any principal's interim posterior, an increase in the informativeness of the agent's signal weakly improves the principal's payoff at the optimal delegation decision.
\end{proposition}

\begin{proof}
First, we establish that the case where a marginal increase in the agent's informativeness can reduce the principal's delegation payoff only happens when the principal was indifferent between delegation and direct action under the original signal. To see this, note that a decrease in delegation payoff results from one of the final posteriors jumping downward at the discontinuity. Therefore, before the discontinuity, the two final posteriors must be co-linear with the interim posterior, making the principal indifferent between delegation or direct action. In this case, it is weakly optimal for the principal to take direct action at $\mu_{s_1}$ before and after the change in the agent's informativeness. Since the payoff of direct action does not change with the agent's signal, the principal's optimal payoff remains constant. 

In all remaining cases, a marginal increase in the agent's informativeness weakly improves the principal's delegation payoff. Since the non-delegation payoff is unaffected, the payoff at the optimal delegation decision also weakly improves. 
\end{proof}

Taken together, Propositions~\ref{prop:cs_q} and \ref{prop:cs_q2} address the question: is more information from the agent always better? The answer depends on whether the tool of delegation is available to the principal. If the principal is free to take her own actions, then having access to a more informed agent is always weakly beneficial. If the principal is required to delegate, however, an improvement in agent's informativeness does not guarantee a monotonic improvement in the payoff of the principal.

\subsection{Degree of Preference Misalignment}
The principal's delegation decision trades off the gain from the agent's private signal and the loss from preference misalignment. This section studies how the principal's payoffs change when the agent becomes less aligned with the principal. 

\begin{definition}
    Holding the principal's preferences fixed, we say that an agent is \emph{more misaligned} if the disagreement interval induced by the preferences of the new agent strict contains that of the original agent.
\end{definition}

\begin{figure}[htb]
  \centering
  \subfigure{  \begin{tikzpicture}[scale=0.8, transform shape]
    \pgfplotsset{compat=1.17}
    \begin{axis}[
      xlabel={$\Pr(\Theta=1)$},
      ylabel={$U$},
      xmin=-0.1, xmax=1,
      axis lines=middle,
      samples=100,
      xtick=\empty,
      ytick=\empty,
      xlabel style={anchor=west},
      ylabel style={anchor=east},
      clip=false,
      ]
      \addplot[black, thick, domain=0.75:1] {x};
      \addplot[black, dashed, domain=0:0.75] {x};
      \addplot[black, thick, domain=0:0.75] {1 - x};
      \addplot[black, dashed, domain=0.75:1] {1 - x};
      \addplot[red, thick, domain=0.75:0.85] {1 - x};
      \addplot[blue, thick, domain=0.75:0.85] {x};

      \coordinate (A) at (0.5, 0.5);
      \coordinate (B) at (0.75, 0.75);
      \coordinate (C) at (0.5, 0);
      \coordinate (D) at (0.75, 0);
      \coordinate (E) at (0.85, 0);

      \node[draw,circle,inner sep=1pt,fill=black] at (C) {};
      \node[draw,circle,inner sep=1pt,fill=black] at (D) {};
      \node[draw,circle,inner sep=1pt,fill=red] at (E) {};

      \node[below left] at (0, 0) {$0$};
      \node[below] at (1, 0) {1};
      \draw[dashed] (A) -- (0.5, 0);
      \draw[dashed, blue] (B) -- (0.75, 0);

      \draw[dashed, red] (E) -- (0.85, 0.85);
      \draw[<->] (0.5,0.05) -- (0.75,0.05) node[midway, fill=white] {$I$};
      \draw[<->] (0.5,-0.05) -- (0.85,-0.05) node[midway, fill=white] {$I'$};
    
    \end{axis}
  \end{tikzpicture}}
  \caption{A more misaligned agent. The black plus the blue sections represent the original delegation envelope $V_D$. The black plus the red sections represent the new delegation envelope $V'_D$.}\label{fig:cs:misalign}
\end{figure}
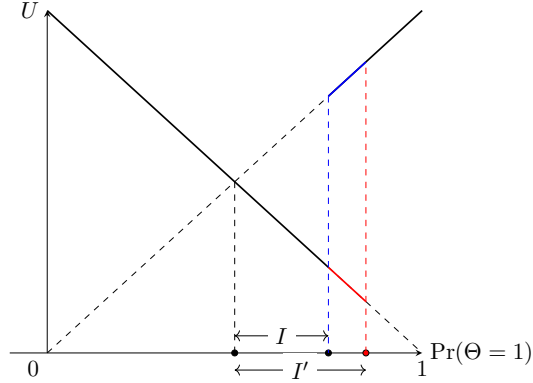

\begin{proposition}\label{prop:cs_l}
For any principal's interim posterior and any agent's signal, a more misaligned agent weakly decreases the principal's delegation payoff and payoff at the optimal delegation decision. 
\end{proposition}

\begin{proof}
    Suppose that the original disagreement interval is $I$ and that it expands to $I' \supset I$ under the new agent. Denote the principal's original delegation envelope $V_D$ and new delegation envelope $V'_D$.

    We consider the principal's delegation payoff. First, consider the $\mu_{s_1}$ and $S_2$ such that both $\mu_{s_10}$ and $\mu_{s_11}$ lie outside of the interval $I'\setminus I$. Observe that $V_D(x) = V'_D(x)$ for all $x \notin I'\setminus I$. Therefore, $V_D(\mu_{s_1s_2})=V'_D(\mu_{s_1s_2})$. This implies that the delegation payoff $\E_{s_2}[V'_D(\mu_{s_1s_2})]$, a linear combination of $V'_D(\mu_{s_10})$ and $V'_D(\mu_{s_11})$, also stays the same as that under the original agent.

    Then, we consider the $\mu_{s_1}$ and $S_2$ such that at least one of the $\mu_{s_10}$ and $\mu_{s_11}$ lies in $I'\setminus I$. Observe that $V'_D(x) < V_D(x)$ for all $x \in I'\setminus I$. Therefore, $V'_D(\mu_{s_1s_2})< V_D(\mu_{s_1s_2})$ for at least one of realizations of $S_2$, and stays the same for the other (or neither) realization. The delegation payoff $\E_{s_2}[V'_D(\mu_{s_1s_2})]$, a linear combination of $V'_D(\mu_{s_10})$ and $V'_D(\mu_{s_11})$, is (strictly) lower than that under the original agent.

    We have proved that the principal's delegation payoff weakly decreases under a more misaligned agent. We note that the principal's nondelegation payoff is the same regardless of the agent available to her. Therefore, the principal's payoff at the optimal delegation decision, which is the maximum of the two payoffs, also weakly decreases under a more misaligned agent. 
\end{proof}

\begin{corollary}\label{corr:cs_l}
    For some principal's interim posterior and some agent's signal, a more misaligned agent strictly decreases the principal's delegation payoff and payoff at the optimal delegation decision. 
\end{corollary}

Proposition~\ref{prop:cs_l} and Corollary \ref{corr:cs_l} also provide clues on the change of the principal's delegation behavior when the agent becomes more misaligned. If the principal does not delegate before at a given interim posterior, she would not delegate when the agent becomes more misaligned. If the principal delegates before at a given interim posterior, she may no longer delegate to a more misaligned agent.

So far, we have considered the change in preference misalignment induced by a change in the agent's preferences. It is also possible that a greater preference misalignment is induced by a change in the principal's preference parameters. In Appendix~\ref{app:cs:principal}, we explore comparative statics results on the principal's payoffs under such cases.

\section{Information Design}

The principal not only decides whether to give decision authority to the agent but also designs the informational structure of the public signal. In choosing between public signals, she must take into account the optimal delegation decision in the second stage after the public signal is realized. 

We are interested in the optimal design of the public signal conditional on making the optimal delegation decision in the second stage. If the signal space is unconstrained, it is trivial that the principal would simply choose the state-revealing signal and never delegate. After we impose constraints on the maximal Blackwell informativeness in the signal space, this question becomes nontrivial: will the principal always choose the most informative signal within those constraints?  We show that the answer is no, even if the principal has the option to use the signal by herself. 

We first introduce a few more notations that would facilitate the analysis. Recall that the red diamond point in Figure~\ref{fig:valuable} represents the principal's \emph{ex-ante delegation payoff}, $\E_{s_2}[V_D(\mu_{s_1s_2})]$, at the interim posterior $\mu_{s_1}$. Holding fixed the agent's signal and player's preferences, this value is a function of the principal's interim posterior, $\mu_{s_1}$. Therefore, we can plot $\E_{s_2}[V_D(\mu_{s_1s_2})]$ as a function of $\mu_{s_1}$ for any fixed agent's signal and player's preferences. 

\begin{figure}[htb]
  \centering
  \subfigure{\begin{tikzpicture}[scale=0.8, transform shape]
  \begin{axis}[
    xlabel={$\mu_{s_1}$},
    ylabel={$U$},
    xmin=-0.1, xmax=1,
    ymin=-0.1, ymax=1,
    axis lines=middle,
    samples=100,
    xtick=\empty,
    ytick=\empty,
    xlabel style={anchor=west},
    ylabel style={anchor=east},
    clip=false,
    ]
    \addplot[black, thick, domain=0.9:1] {x};
    \addplot[black, dashed, domain=0.75:1] {x};
    \addplot[black, thick, domain=0.45:0.9] {0.8};
    \addplot[black, thick, domain=0:0.45] {1 - x};
    \addplot[black, dashed, domain=0:0.75] {1 - x};
    
    \coordinate (A) at (0.5, 0.5);
    \coordinate (B) at (0.75, 0.75);
    \coordinate (C) at (0.5, 0);
    \coordinate (D) at (0.75, 0);
    
\node[draw,diamond,inner sep=1pt,fill=red] at (0.45,0.8) {};
 \node[draw,circle,inner sep=1pt,fill=red] at (0.75,0.75) {};
 \node[draw,circle,inner sep=1pt,fill=red] at (0.155,0.845) {};
 \draw [thick, red, dashed](0.75,0.75)--(0.155,0.845);
    \node[below left] at (0, 0) {$0$};
    \node[below] at (1, 0) {1};

  \end{axis}
\end{tikzpicture}}
  \caption{Principal's ex-ante delegation payoff as a function of the interim posterior}\label{fig:exante_D}
\end{figure}
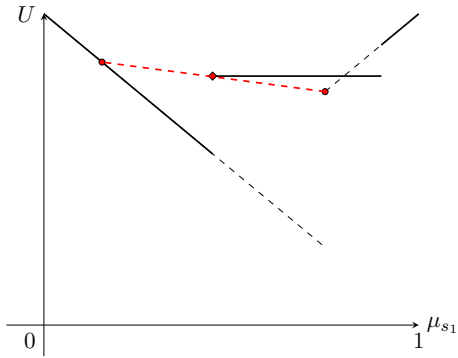

In Figure \ref{fig:exante_D}, principal's delegation envelope ($V_D$) is plotted as dashed lines and her ex-ante delegation payoff ($\E_{s_2}[V_D(\mu_{s_1s_2})]$) is plotted as solid lines. Each point on the solid line is the probability weighted average of two points on the dashed line.\footnote{\enskip The weights are the probabilities of $S_2 = 0$ and $S_2 = 1$ conditional on $s_1$.} The red points provide one such example. When the principal's preference and the agent's signal are symmetric ($r_{00} = r_{11}$, $r_{01} = r_{10}$, and $q_0=q_1=q$), averaging produces a constant payoff in the intermediate section. This special case is proved in Appendix~\ref{app:special_q}. In general, the function is a piecewise linear function with three sections: the left and right sections follow the delegation envelope, while the middle section can be flat, upward, or downward sloping. 

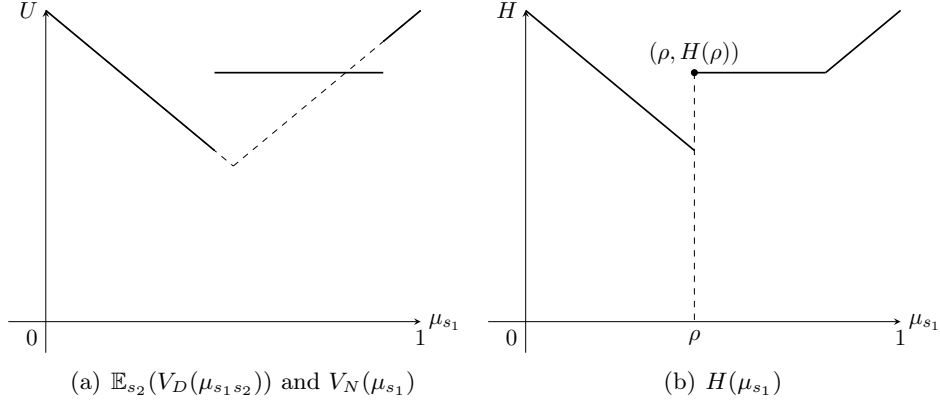
\begin{figure}[htb]
  \centering
  \subfigure[$\E_{s_2}(V_D(\mu_{s_1s_2}))$ and $V_N(\mu_{s_1})$\label{fig:exanteD_ND}]
  {\begin{tikzpicture}[scale=0.8, transform shape]
  \begin{axis}[
    xlabel={$\mu_{s_1}$},
    ylabel={$U$},
    xmin=-0.1, xmax=1,
    ymin=-0.1, ymax=1,
    axis lines=middle,
    samples=100,
    xtick=\empty,
    ytick=\empty,
    xlabel style={anchor=west},
    ylabel style={anchor=east},
    clip=false,
    ]
    \addplot[black, thick, domain=0.9:1] {x};
    \addplot[black, dashed, domain=0.5:1] {x};
    \addplot[black, thick, domain=0.45:0.9] {0.8};
    \addplot[black, thick, domain=0:0.45] {1 - x};
    \addplot[black, dashed, domain=0:0.5] {1 - x};
    
    \coordinate (A) at (0.5, 0.5);
    \coordinate (B) at (0.75, 0.75);
    \coordinate (C) at (0.5, 0);
    \coordinate (D) at (0.75, 0);
    

    \node[below left] at (0, 0) {$0$};
    \node[below] at (1, 0) {1};

  \end{axis}
\end{tikzpicture}}
\subfigure[$H(\mu_{s_1})$\label{fig:H_ND}]{\begin{tikzpicture}[scale=0.8, transform shape]
  \begin{axis}[
    xlabel={$\mu_{s_1}$},
    ylabel={$H$},
    xmin=-0.1, xmax=1,
    ymin=-0.1, ymax=1,
    axis lines=middle,
    samples=100,
    xtick=\empty,
    ytick=\empty,
    xlabel style={anchor=west},
    ylabel style={anchor=east},
    clip=false,
    ]
    \addplot[black, thick, domain=0.8:1] {x};
    \addplot[black, thick, domain=0.45:0.8] {0.8};
    \addplot[black, thick, domain=0:0.45] {1 - x};

    \coordinate (A) at (0.45, 0.8);
    \coordinate (B) at (0.2, 0.8);
    \coordinate (C) at (0.45, 0);
    \coordinate (D) at (0.2, 0);
    
    \node[draw,circle,inner sep=1pt,fill=black] at (A) {};

    \node[below left] at (0, 0) {$0$};
    \node[below] at (1, 0) {1};
    \draw[dashed] (A) -- (C);
    \node[below] at (C) {$\rho$};
    \node[above] at (A) {$(\rho, H(\rho))$};

  \end{axis}
\end{tikzpicture}}
  \caption{Principal's optimal delegation-stage payoff}\label{fig:H}
\end{figure}

The value of this function represents the principal's payoff if she were to delegate at a given interim posterior $\mu_{s_1}$. However, this is only one part of the equation in what the principal can obtain at the delegation stage. She also has the choice of taking an action directly according to her interim posterior. Therefore, principal's \emph{optimal delegation-stage payoff} is the maximum between the ex-ante delegation payoff and the payoff on the non-delegation envelope. Mathematically, we define:
    \[H(\mu_{s_1})\coloneq \max\{\E_{s_2}[V_D(\mu_{s_1s_2})],V_N(\mu_{s_1})\}\]
Figure \ref{fig:exanteD_ND} and \ref{fig:H_ND} plot the principal's ex-ante delegation payoff, the non-delegation envelope, and the resulting $H(\mu_{s_1})$ as the maximum of the two. It is also possible to have the discontinuity past the principal's cutoff point $\bar{\mu}_P$, which results in $H(\mu_{s_1})$ as plotted in Appendix \ref{app:plot}, but all analyses below still go through in such cases.

One important point in the analysis is the point of discontinuity, which we label as $(\rho,H(\rho))$. Since we made the assumption that the agent takes the principal-preferred action when indifferent, $H$ is upper semicontinuous at $\rho$. In words, $\rho$ is the interim posterior such that, upon receiving another principal-preferred signal realization, the agent's final posterior equals $\bar{\mu}_A$ (and thus is indifferent between two actions). Algebraically, 
\begin{align*}
    \rho = \frac{(1-q_0)\bar{\mu}_A}{(1-q_0)\bar{\mu}_A+q_1(1-\bar{\mu}_A)} \text{   if   } \bar{\mu}_A>\bar{\mu}_P \\
    \rho = \frac{ q_0 \bar{\mu}_A}{ q_0 \bar{\mu}_A+(1-q_1)(1-\bar{\mu}_A)} \text{   if   } \bar{\mu}_A<\bar{\mu}_P 
\end{align*}

The principal's information design problem is thus choosing the public signal that, given prior $\mu$ and Blackwell constraint $\lambda$, generates two interim posteriors $\mu_{0}$ and $\mu_{1}$ such that $H(\mu_{0})$ and $H(\mu_{1})$ give the maximal expected payoff. Recall that we denote the posteriors generated by the maximally informative signal subject to Blackwell constraint $\lambda$ as $\mu_0^\lambda$ and $\mu_1^\lambda$. We are now ready to state our main proposition for the optimal information design.

\begin{proposition}[Optimal information design]\label{prop:infodesign}
    Given a prior $\mu$ and a Blackwell constraint $\lambda$, the principal chooses the most informative public signal if and only if there exists a weakly convex and continuous function $c$ such that the points $(\mu_0^\lambda,H(\mu_0^\lambda))$, $(\mu_1^\lambda,H(\mu_1^\lambda))$, and $(\rho,H(\rho))$ are in the graph of $c$.\footnote{\enskip For the if direction, we need to assume that the principal chooses the most informative signal when indifferent.} If this is the case, we say that these three points can be ``convexified''.
    
   If they cannot be convexified,
    \begin{enumerate}
        \item If $\mu>\rho$, $\mu_1^* = \mu_1^\lambda$ and $\mu_0^* = \rho$. The optimal signal generates the high posterior at the Blackwell constraint and the low posterior at $\rho$. 
        \item If $\mu<\rho$, $\mu_0^* = \mu_0^\lambda$ and $\mu_1^* = \rho$. The optimal signal generates the low posterior at the Blackwell constraint and the high posterior at $\rho$. 
        \item If $\mu=\rho$, $\mu_1^*=\mu_0^*=\mu$. The optimal signal is completely uninformative. 
    \end{enumerate}
\end{proposition}

\begin{figure}[htp]
  \centering
  \subfigure[Can be convexified\label{fig:infodesign1}]{\begin{tikzpicture}[scale=0.8, transform shape]
  \begin{axis}[
    xlabel={$\mu_{s_1}$},
    ylabel={$H$},
    xmin=-0.1, xmax=1,
    ymin=-0.1, ymax=1,
    axis lines=middle,
    samples=100,
    xtick=\empty,
    ytick=\empty,
    xlabel style={anchor=west},
    ylabel style={anchor=east},
    clip=false,
    ]
    \addplot[black, thick, domain=0.8:1] {x};
    \addplot[black, thick, domain=0.45:0.8] {0.8};
    \addplot[black, thick, domain=0:0.45] {1 - x};

    \coordinate (A) at (0.45, 0.8);
    \coordinate (C) at (0.45, 0);
    \coordinate (M) at (0.6, 0.8);
    \coordinate (M0) at (0.1, 0.9);
    \coordinate (M1) at (0.9, 0.9);
    \coordinate (MX) at (0.6, 0);
    \coordinate (M0X) at (0.1, 0);
    \coordinate (M1X) at (0.9, 0);    
    
    \node[draw,circle,inner sep=1pt,fill=black] at (A) {};
    \node[draw,diamond,inner sep=1.5pt,fill=black] at (M) {};
    \node[draw,circle,inner sep=1pt,fill=black] at (M0) {};
    \node[draw,circle,inner sep=1pt,fill=black] at (M1) {};
    
    \node[below left] at (0, 0) {$0$};
    \node[below] at (1, 0) {1};
    \draw[dashed] (A) -- (C);
    \draw[dashed] (M) -- (MX);
    \draw[dashed] (M0) -- (M0X);
    \draw[dashed] (M1) -- (M1X);
    \node[below] at (C) {$\rho$};
    \node[above] at (A) {$(\rho, H(\rho))$};
    \node[above right] at (M0) {$(\mu_0, H(\mu_0))$};
    \node[above left] at (M1) {$(\mu_1, H(\mu_1))$};
    \node[below] at (MX) {$\mu$};
    \node[below] at (M0X) {$\mu_0$};
    \node[below] at (M1X) {$\mu_1$};

    \draw (M0) edge[parabola through={(A)},
    red,thick] (M1);
    
  \end{axis}
\end{tikzpicture}}
  \subfigure[Cannot be convexified, optimal $\mu^*_0 = \rho$\label{fig:infodesign2}]{\begin{tikzpicture}[scale=0.8, transform shape]
  \begin{axis}[
    xlabel={$\mu_{s_1}$},
    ylabel={$H$},
    xmin=-0.1, xmax=1,
    ymin=-0.1, ymax=1,
    axis lines=middle,
    samples=100,
    xtick=\empty,
    ytick=\empty,
    xlabel style={anchor=west},
    ylabel style={anchor=east},
    clip=false,
    ]
    \addplot[black, thick, domain=0.8:1] {x};
    \addplot[black, thick, domain=0.45:0.8] {0.8};
    \addplot[black, thick, domain=0:0.45] {1 - x};

    \coordinate (A) at (0.45, 0.8);
    \coordinate (C) at (0.45, 0);
    \coordinate (M) at (0.6, 0.8);
    \coordinate (M0) at (0.3, 0.7);
    \coordinate (M1) at (0.85, 0.85);
    \coordinate (MX) at (0.6, 0);
    \coordinate (M0X) at (0.3, 0);
    \coordinate (M1X) at (0.85, 0);    
    
    \node[draw,circle,inner sep=1pt,fill=black] at (A) {};
    \node[draw,diamond,inner sep=1.5pt,fill=black] at (M) {};
    \node[draw,circle,inner sep=1pt,fill=black] at (M0) {};
    \node[draw,circle,inner sep=1pt,fill=black] at (M1) {};
    
    \node[below left] at (0, 0) {$0$};
    \node[below] at (1, 0) {1};
    \draw[dashed] (A) -- (C);
    \draw[dashed] (M) -- (MX);
    \draw[dashed] (M0) -- (M0X);
    \draw[dashed] (M1) -- (M1X);
    \node[below] at (C) {$\rho$};
    \node[above] at (A) {$(\rho, H(\rho))$};
    \node[below left] at (M0) {$(\mu_0, H(\mu_0))$};
    \node[below right] at (M1) {$(\mu_1, H(\mu_1))$};
    \node[below] at (MX) {$\mu$};
    \node[below] at (M0X) {$\mu_0$};
    \node[below] at (M1X) {$\mu_1$};

    \draw (M0) edge[parabola through={(A)},
    red,thick,dashed] (M1);
    \draw (A) edge[blue,thick] (M1);
    
  \end{axis}
\end{tikzpicture}}
  \subfigure[Can be convexified\label{fig:infodesign3}]{\begin{tikzpicture}[scale=0.8, transform shape]
  \begin{axis}[
    xlabel={$\mu_{s_1}$},
    ylabel={$H$},
    xmin=-0.1, xmax=1,
    ymin=-0.1, ymax=1,
    axis lines=middle,
    samples=100,
    xtick=\empty,
    ytick=\empty,
    xlabel style={anchor=west},
    ylabel style={anchor=east},
    clip=false,
    ]
    \addplot[black, thick, domain=0.8:1] {x};
    \addplot[black, thick, domain=0.45:0.8] {0.8};
    \addplot[black, thick, domain=0:0.45] {1 - x};

    \coordinate (A) at (0.45, 0.8);
    \coordinate (C) at (0.45, 0);
    \coordinate (M) at (0.4, 0.6);
    \coordinate (M0) at (0.05, 0.95);
    \coordinate (M1) at (0.8, 0.8);
    \coordinate (MX) at (0.4, 0);
    \coordinate (M0X) at (0.05, 0);
    \coordinate (M1X) at (0.8, 0);    
    
    \node[draw,circle,inner sep=1pt,fill=black] at (A) {};
    \node[draw,diamond,inner sep=1.5pt,fill=black] at (M) {};
    \node[draw,circle,inner sep=1pt,fill=black] at (M0) {};
    \node[draw,circle,inner sep=1pt,fill=black] at (M1) {};
    
    \node[below left] at (0, 0) {$0$};
    \node[below] at (1, 0) {1};
    \draw[dashed] (A) -- (C);
    \draw[dashed] (M) -- (MX);
    \draw[dashed] (M0) -- (M0X);
    \draw[dashed] (M1) -- (M1X);
    \node[below] at (C) {$\rho$};
    \node[above] at (A) {$(\rho, H(\rho))$};
    \node[above right] at (M0) {$(\mu_0, H(\mu_0))$};
    \node[below right] at (M1) {$(\mu_1, H(\mu_1))$};
    \node[below] at (MX) {$\mu$};
    \node[below] at (M0X) {$\mu_0$};
    \node[below] at (M1X) {$\mu_1$};

    \draw (M0) edge[parabola through={(A)},
    red,thick] (M1);
    
  \end{axis}
\end{tikzpicture}}
  \subfigure[Cannot be convexified, optimal $\mu^*_1 = \rho$\label{fig:infodesign4}]{\begin{tikzpicture}[scale=0.8, transform shape]
  \begin{axis}[
    xlabel={$\mu_{s_1}$},
    ylabel={$H$},
    xmin=-0.1, xmax=1,
    ymin=-0.1, ymax=1,
    axis lines=middle,
    samples=100,
    xtick=\empty,
    ytick=\empty,
    xlabel style={anchor=west},
    ylabel style={anchor=east},
    clip=false,
    ]
    \addplot[black, thick, domain=0.8:1] {x};
    \addplot[black, thick, domain=0.45:0.8] {0.8};
    \addplot[black, thick, domain=0:0.45] {1 - x};

    \coordinate (A) at (0.45, 0.8);
    \coordinate (C) at (0.45, 0);
    \coordinate (M) at (0.4, 0.6);
    \coordinate (M0) at (0.3, 0.7);
    \coordinate (M1) at (0.6, 0.8);
    \coordinate (MX) at (0.4, 0);
    \coordinate (M0X) at (0.3, 0);
    \coordinate (M1X) at (0.6, 0);    
    
    \node[draw,circle,inner sep=1pt,fill=black] at (A) {};
    \node[draw,diamond,inner sep=1.5pt,fill=black] at (M) {};
    \node[draw,circle,inner sep=1pt,fill=black] at (M0) {};
    \node[draw,circle,inner sep=1pt,fill=black] at (M1) {};
    
    \node[below left] at (0, 0) {$0$};
    \node[below] at (1, 0) {1};
    \draw[dashed] (A) -- (C);
    \draw[dashed] (M) -- (MX);
    \draw[dashed] (M0) -- (M0X);
    \draw[dashed] (M1) -- (M1X);
    \node[below] at (C) {$\rho$};
    \node[above] at (A) {$(\rho, H(\rho))$};
    \node[below left] at (M0) {$(\mu_0, H(\mu_0))$};
    \node[below right] at (M1) {$(\mu_1, H(\mu_1))$};
    \node[below] at (MX) {$\mu$};
    \node[below] at (M0X) {$\mu_0$};
    \node[below] at (M1X) {$\mu_1$};

    \draw (M0) edge[parabola through={(A)},
    red,thick,dashed] (M1);
    \draw (A) edge[blue,thick] (M0);
  \end{axis}
\end{tikzpicture}}
  \subfigure[Trivially convexified\label{fig:infodesign5}]{\begin{tikzpicture}[scale=0.8, transform shape]
  \begin{axis}[
    xlabel={$\mu_{s_1}$},
    ylabel={$H$},
    xmin=-0.1, xmax=1,
    ymin=-0.1, ymax=1,
    axis lines=middle,
    samples=100,
    xtick=\empty,
    ytick=\empty,
    xlabel style={anchor=west},
    ylabel style={anchor=east},
    clip=false,
    ]
    \addplot[black, thick, domain=0.8:1] {x};
    \addplot[black, thick, domain=0.45:0.8] {0.8};
    \addplot[black, thick, domain=0:0.45] {1 - x};

    \coordinate (A) at (0.45, 0.8);
    \coordinate (C) at (0.45, 0);
    \coordinate (M) at (0.7, 0.8);
    \coordinate (M0) at (0.55, 0.8);
    \coordinate (M1) at (0.9, 0.9);
    \coordinate (MX) at (0.7, 0);
    \coordinate (M0X) at (0.55, 0);
    \coordinate (M1X) at (0.9, 0);    
    
    \node[draw,circle,inner sep=1pt,fill=black] at (A) {};
    \node[draw,diamond,inner sep=1.5pt,fill=black] at (M) {};
    \node[draw,circle,inner sep=1pt,fill=black] at (M0) {};
    \node[draw,circle,inner sep=1pt,fill=black] at (M1) {};
    
    \node[below left] at (0, 0) {$0$};
    \node[below] at (1, 0) {1};
    \draw[dashed] (A) -- (C);
    \draw[dashed] (M) -- (MX);
    \draw[dashed] (M0) -- (M0X);
    \draw[dashed] (M1) -- (M1X);
    \node[below] at (C) {$\rho$};
    \node[above] at (A) {$(\rho, H(\rho))$};
    \node[below] at (M0) {$(\mu_0, H(\mu_0))$};
    \node[right] at (M1) {$(\mu_1, H(\mu_1))$};
    \node[below] at (MX) {$\mu$};
    \node[below] at (M0X) {$\mu_0$};
    \node[below] at (M1X) {$\mu_1$};

    \draw (A) edge[parabola through={(M0)},
    red,thick] (M1);
    
  \end{axis}
\end{tikzpicture}}
  \subfigure[Trivially convexified\label{fig:infodesign6}]{\begin{tikzpicture}[scale=0.8, transform shape]
  \begin{axis}[
    xlabel={$\mu_{s_1}$},
    ylabel={$H$},
    xmin=-0.1, xmax=1,
    ymin=-0.1, ymax=1,
    axis lines=middle,
    samples=100,
    xtick=\empty,
    ytick=\empty,
    xlabel style={anchor=west},
    ylabel style={anchor=east},
    clip=false,
    ]
    \addplot[black, thick, domain=0.8:1] {x};
    \addplot[black, thick, domain=0.45:0.8] {0.8};
    \addplot[black, thick, domain=0:0.45] {1 - x};

    \coordinate (A) at (0.45, 0.8);
    \coordinate (C) at (0.45, 0);
    \coordinate (M) at (0.2, 0.8);
    \coordinate (M0) at (0.05, 0.95);
    \coordinate (M1) at (0.35, 0.65);
    \coordinate (MX) at (0.2, 0);
    \coordinate (M0X) at (0.05, 0);
    \coordinate (M1X) at (0.35, 0);    
    
    \node[draw,circle,inner sep=1pt,fill=black] at (A) {};
    \node[draw,diamond,inner sep=1.5pt,fill=black] at (M) {};
    \node[draw,circle,inner sep=1pt,fill=black] at (M0) {};
    \node[draw,circle,inner sep=1pt,fill=black] at (M1) {};
    
    \node[below left] at (0, 0) {$0$};
    \node[below] at (1, 0) {1};
    \draw[dashed] (A) -- (C);
    \draw[dashed] (M) -- (MX);
    \draw[dashed] (M0) -- (M0X);
    \draw[dashed] (M1) -- (M1X);
    \node[below] at (C) {$\rho$};
    \node[above] at (A) {$(\rho, H(\rho))$};
    \node[above right] at (M0) {$(\mu_0, H(\mu_0))$};
    \node[right] at (M1) {$(\mu_1, H(\mu_1))$};
    \node[below] at (MX) {$\mu$};
    \node[below] at (M0X) {$\mu_0$};
    \node[below] at (M1X) {$\mu_1$};

    \draw (M0) edge[parabola through={(M1)},
    red,thick] (A);
    
  \end{axis}
\end{tikzpicture}}
  \caption{Optimal information design}\label{fig:infodesign}
\end{figure}
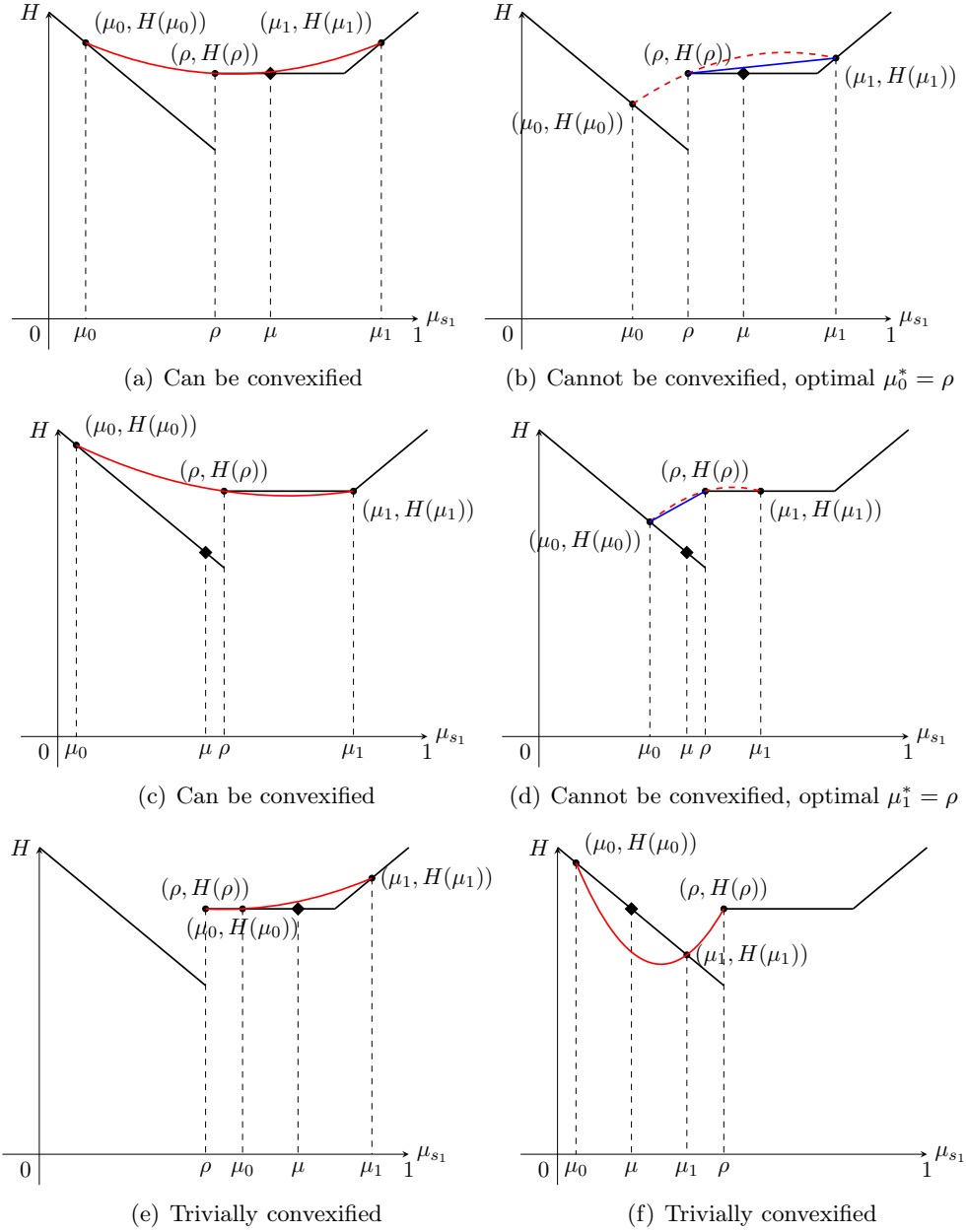

\begin{proof}
First, if the posteriors of the maximally informative signal lie on the same side of the discontinuity at $\rho$, then it is trivial that they can be convexified with $(\rho,H(\rho))$ as one of the end points (Figure~\ref{fig:infodesign5}, \ref{fig:infodesign6}). In this case, the maximal signal is at least weakly optimal since expanding posteriors on a weakly convex function generates a weakly higher expected payoff. 

Next, we consider the case when the two posteriors lie on different sides of the discontinuity at $\rho$. Observe that in this case the principal's expected payoff always weakly increases when the two posteriors marginally move away from the prior (when the signal becomes more Blackwell informative). Therefore, the maximal signal gives the highest expected payoff among all the signals that generate posteriors on different sides of $(\rho,H(\rho))$. What we need to compare now is the payoff of the maximal signal with the payoff of having one of the posteriors stop at $\rho$. 

Suppose that the three points can be convexified, with $(\rho,H(\rho))$ in the middle (Figure~\ref{fig:infodesign1}, \ref{fig:infodesign3}). The convexity of the connecting function ensures that the line segment between the two posteriors lies weakly above the line segment connecting either of the posterior with $(\rho,H(\rho))$. Therefore, no matter where the prior, the expected payoff generated by the maximal signal exceeds that generated by placing one of the posteriors at $\rho$. Suppose instead that there exists no weakly convex and continuous function connecting the three points. In the case of three points, it has to be the case that they can be connected by a strictly concave function. Then, the line segment between the two posteriors lies strictly below the line segment connecting one of the posteriors with $(\rho,H(\rho))$. This proves that restricting one posterior to $\rho$ produces strictly higher expected payoff than the maximal signal. How about moving the posterior even closer to $\mu$? This is not optimal since the section of the curve between $\rho$ and $\mu$ is now weakly convex so contracting the posteriors gives a weakly lower payoff. 
\end{proof}

Why is the point $(\rho,H(\rho))$ the optimal posterior when the principal does not give out the most informative signal? As argued before, placing one of the interim posteriors at $\rho$ would make the agent indifferent between the two actions if he receives a principal-preferred realization of the private signal. Therefore, he would take the principal-preferred action out of indifference. If instead, the agent strictly prefers to take the principal-preferred action, then the principal can increase the probability of the principal-preferred realization in the agent-preferred state and still ensures that agent takes the same action. 

For example, suppose that the principal preferred action is $1$. If the agent strictly prefers to take action $1$ when observing $(S_1=1, S_2=1)$, then the principal could slightly decrease $\Pr(S_1=0\mid \Theta=0)$ and increase $\Pr(S_1=1\mid \Theta=0)$ to lower the interim posterior $\mu_{11}$ back to $\rho$, while still ensuring that the agent takes action $1$ when observing $(S_1=1, S_2=1)$. The principal has thus constructed a signal with higher chance of realizing $S_1=1$ while leaving the agent's optimal action given the signal realization unchanged, and thus strictly increased her payoff. This is a generalization of the indifference result in \citet{kamenica_bayesian_2011} applied to our setting where the receiver can receive an additional signal.

\subsection{Features of the Optimal Information Design}
This section highlights features of the optimal information design that may inform the design of real-life algorithms that interact with misaligned agents.

\paragraph{Maximizing prediction accuracy in the presence of a biased agent.} One implication of the optimal information design is that whether one should provide the most informative public signal depends on what the maximally informative signal can achieve. For a given prior, if the maximal signal is sufficiently uninformative, then there is no harm in providing it because: 1) the principal can use it in the delegation stage; and 2) even if the principal decides to delegate, the public signal is not strong enough to sway the agent's decision. If the maximal signal can generate interim posteriors sufficiently confident (near the end points), then it is strictly optimal to maximize informativeness because the principal and the agent tend to agree near the extreme beliefs. However, if the maximal signal only provides moderately confident interim posteriors, it is usually optimal for the principal to delegate in order to obtain more information. Given that the principal will delegate in the second stage, a ``skewed'' signal designed to persuade the agent can do better than providing the maximally informative signal. 

\paragraph{One-sided information.}
In all but the the edge case ($\mu=\rho$), the optimal information structure maximizes the signal strength on one side. Specifically, when the prior is on the right (left) of the discontinuity point $\rho$, the optimal signal places the posterior at the Blackwell constraint on the right (left) side. This implies different investment decisions between lowering an algorithm's false negative rate ($1-p_1$) versus false positive rate ($1-p_0$) when we care about quality of the the final decisions rather than that of the predictions. For example, if $\mu > \rho$, the optimal signal maximizes the high posterior $\mu_1$ while potentially restricting the low posterior $\mu_0$. This implies that the optimal signal features a very low false positive rate and a moderate false negative rate.

\section{Policy Restrictions and Decision Quality}
In our model, the principal can utilize the tools of persuasion and delegation to mitigate the agent's preference misalignment. 
First, as she designs the information structure of the public signal, she can take into account the possibility of passing the information and the decision to the misaligned agent.
Second, upon each signal realization, she can decide whether to delegate the decision to the agent or to act directly based on the current information. 

In real-life applications, however, the principal may have limited access to these tools due to legal, institutional, or moral constraints. For example, Article 14 of the EU AI act requires that high-risk AI systems must be designed such that they can be ``effectively overseen by natural persons'' \citep{EU_Artificial_Intelligence_Act_2021}. In the context of our model, it implies that the principal is restricted to always delegate.
In other applications, the algorithm may be required to be ``truthful,'' meaning it simply provides the most informative signal regardless of the human it interacts with. Indeed, in most traditional applications of decision-aid algorithms, neither of these strategic tools were available. The algorithm is no more than a prediction tool with no potential to persuade nor any authority to take actions by itself. 

Indeed, these policies, such as designing maximally informative algorithms and obtaining additional information from human agents, improves decision quality when there is no preference misalignment. However, in situations where there is an interaction between the potential biases of the agent and information provision, our analysis show that these naive policies may in fact be welfare worsening for the principal. 
This is formalized in the following propositions.

\begin{proposition}\label{prop:welfare1}
    If $\bar{\mu}_P \neq \bar{\mu}_A$ and $q_0, q_1 <1$, there exists interim posterior $\mu_{s_1}$ at which delegating to the agent is strictly worse than taking direct action.
\end{proposition}

\begin{proof}
    The principal benefits from the option to not delegate when at the interim posterior, the ex-ante payoff of delegation is strictly lower than the non-delegation envelope. In Figure~\ref{fig:exanteD_ND}, it is the part when the solid line (ex-ante payoff of delegation) is strictly lower than the dashed line (non-delegation envelope). This part of sub-optimal delegation is generated by averaging between two final posteriors, one of which falls into the disagreement interval. Unless $\bar{\mu}_P=\bar{\mu}_A$, the disagreement interval exists. If $q_0, q_1 <1$, the agent's final posterior can fall into the disagreement interval.
\end{proof}

\begin{proposition}\label{prop:welfare2}
    If $\bar{\mu}_P \neq \bar{\mu}_A$ and $q_0, q_1 <1$, there exists prior $\mu$ and Blackwell constraint $\lambda$ such that providing no algorithm is strictly better than providing the maximally informative algorithm. 
\end{proposition}

\begin{proof}
    Providing no (or a completely uninformative) algorithm can be better off than providing the maximal algorithm whenever there is a discontinuity in the optimal delegation-stage payoff and the principal's prior lies in the flat region of $H$ (as in Figure~\ref{fig:infodesign2}). This discontinuity is a product of the agent's final posteriors having a discontinuous jump on the delegation envelope $V_D$. Unless $\bar{\mu}_P=\bar{\mu}_A$, the discontinuity in $V_D$ exists. If $q_0, q_1 <1$, the agent's final posteriors can fall near the discontinuity and create non-convexity in payoffs. In that case, the principal may be better off providing no algorithm and enjoying the delegation payoff at the prior.
\end{proof}

Unless the agent is perfectly aligned or one of his signal realizations is state-revealing, Propositions~\ref{prop:welfare1} and \ref{prop:welfare2} imply that there exist cases where it is strictly beneficial to remove the agent from the decision-making process or to provide no algorithmic assistance to the agent. These theoretical results may explain the empirical puzzle of why the introduction of highly predictive decision-aid algorithms often leads to underperforming human-machine collaborations. For example, \citet{jacobs_how_2021} find that in the medical domain, algorithmic aids did not enhance clinicians' accuracy in choosing antidepressants, with both aided and unaided clinicians underperforming compared to standalone machine-learning systems. Likewise, \citet{imai_experimental_2021} and \citet{stevenson_algorithmic_2022} observe in the judicial context that risk assessment tools barely influenced judges' decisions on pretrial detention and sentencing, failing to notably improve public safety or lower incarceration rates. While these findings may seem puzzling, this paper suggests that the underperformance of human-machine collaborations is not only understandable, but even expected, if no measures are taken to mitigate preference misalignment between algorithm and expert. 

Note that possible misalignment of the agent includes not only any explicit racial or gender biases but also common behavioral biases such as risk aversion, complexity aversion, or preference for inaction. Additionally, the agent must be a Bayesian agent who correctly updates his beliefs upon receiving multiple sources of information. We argue that these conditions are hard to satisfy in real-life human agents. In that case, implementing naive policies that are seemingly conducive to better decision making may in fact worsen decision quality and incur welfare losses for society. 

\section{Discussion and Future Work}
In this paper we consider the joint decision problem of designing prediction algorithms and deciding when to delegate to a privately informed and biased agent. We develop conditions under which delegating to the agent is strictly optimal, and study the change in the principal's payoffs when the agent becomes better informed or more misaligned. We show that given the optimal delegation decision, the optimal algorithm may not be maximally informative. In particular, the optimal algorithm maximizes information about one state while restricting information about the other. In the absence of perfectly aligned
agents and state-revealing signals, we show that there exist cases where it is strictly beneficial to remove the agent from the decision-making process or to provide no algorithmic assistance.

Our characterization highlights that naive policies, such as designing maximally informative algorithms and mandating delegation to human agents, may strictly worsen decision quality for some decision subjects. This implies that even well-intentioned policies aiming to provide more information can produce counter-intuitive results if we fail to account for the strategic interaction between information and biases. Echoing recent works in the Economics and Computation area \citep{xu_decision-aid_2023, mclaughlin_algorithmic_2023}, our results once again demonstrate that good algorithmic predictions do not directly translate to good final decisions. We suggest that the underperformance of human-machine collaborations widely observed in empirical settings can be understood through this theoretical lens. 

We recognize that an important direction for future work is to go beyond theoretical characterizations and estimate human agents’ biases and information from data. Separately identifying agents' biases from their private information can be challenging, as shown in \citet{rambachan2021identifying}. However, it may be possible with quasi-experimental or experimental choice data observed in human-machine collaborative settings. Related to this, it could be worthwhile to empirically discern agents' preference biases (due to different objectives) from updating biases (due to incorrect use of algorithmic predictions). Finally, given that the optimal algorithm interacting with biased agents may withhold information about some states, it would be important to test experimentally whether a strategic algorithm can be detected by users who may then stop trusting the algorithm.

\bibliographystyle{plainnat}
\bibliography{bibliography}

\newpage
\appendix

\section{Justification of Assumption \ref*{ass:payoff}}\label{app:assumption1}
Assumption \ref{ass:payoff} assumes that the principal's payoffs satisfy $r_{00} > r_{01}$ and $r_{11} > r_{10}$. Similarly, the agent's payoffs satisfy $v_{00} > v_{01}$ and $v_{11} > v_{10}$. We will show that this assumption is without loss of generality, since the principal never strictly prefers delegation when this assumption is not satisfied. 

We start with the principal. Suppose that one of the inequality is not satisfied for the principal, say, $r_{00} > r_{01}$ but $r_{11} \leq r_{10}$. In this case, the payoff of taking action $0$ in any state is weakly higher than that of taking action $1$ in the same state. In other words, the principal has a dominant action $0$. Therefore, it is never strictly valuable for the principal to delegate the decision to the agent because the principal would maximize her payoff by taking the dominant action. 

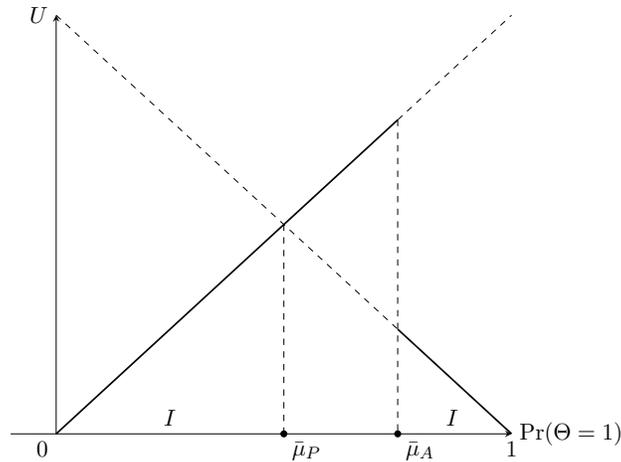
\begin{figure}[htb]
    \centering 
    \scalebox{0.8}{%
    \begin{tikzpicture}
      \pgfplotsset{compat=1.17}
      \begin{axis}[
        xlabel={$\Pr(\Theta=1)$},
        ylabel={$U$},
        xmin=-0.1, xmax=1,
        axis lines=middle,
        samples=100,
        width=0.6\textwidth,
        xtick=\empty,
        ytick=\empty,
        xlabel style={anchor=west},
        ylabel style={anchor=east},
        clip=false,
        ]
        \addplot[black, dashed, domain=0.75:1] {x};
        \addplot[black, thick, domain=0:0.75] {x};
        \addplot[black, dashed, domain=0:0.75] {1 - x};
        \addplot[black, thick, domain=0.75:1] {1 - x};

        \coordinate (A) at (0.5, 0.5);
        \coordinate (B) at (0.75, 0.75);
        \coordinate (C) at (0.5, 0);
        \coordinate (D) at (0.75, 0);
        
        \node[draw,circle,inner sep=1pt,fill=black] at (C) {};
        \node[draw,circle,inner sep=1pt,fill=black] at (D) {};
  
        \node[below left] at (0, 0) {$0$};
        \node[below] at (1, 0) {1};
        \draw[dashed] (A) -- (0.5, 0);
        \node[below right] at (0.5, 0) {$\bar{\mu}_P$};
        \draw[dashed] (B) -- (0.75, 0);
        \node[below right] at (0.75, 0) {$\bar{\mu}_A$};
        \node[above] at (0.25, 0) {$I$};
        \node[above] at (0.87, 0) {$I$};
  
      \end{axis}
    \end{tikzpicture}
    }
    \caption{Principal's payoff envelope of delegation with ``state-mismatching'' preferences}
    \label{app:fig:envelope}
  \end{figure}

Suppose that neither inequalities are satisfied for the principal, i.e., $r_{00} \leq r_{01}$ and $r_{11} \leq r_{10}$. In this case, the principal has a preference for ``mismatching the state.'' In particular, the principal's preferred action switches from $1$ to $0$ as her belief of the probability of state $1$ increases. We further suppose that the agent has the normal ``state-matching'' preferences. Then, the ``disagreement interval (I)'' would in fact be two disjoint intervals at the two ends of the unit interval (as shown in Figure \ref{app:fig:envelope}). The principal's payoff envelope in this case is also depicted in Figure \ref{app:fig:envelope}. It is easy to see that the principal cannot strictly improve her payoff via delegation, since no linear combination on the delegation envelope can be above the non-delegation envelope. 

  We now turn to the assumption on the agent. Same as the principal, the agent would have a dominant action if only one of the inequality for the agent is not satisfied. It is then obvious that the principal cannot have strict improvements via delegation if the agent always take one action.

  If neither inequalities are satisfied for the agent, the problem is symmetric with that of the principal. The ``disagreement'' interval would still be disjoint and the principal's payoff envelope looks like the mirror image of Figure \ref{app:fig:envelope}. It follows that the principal cannot have strict improvements by delegation. 
  
  Finally, if none of the four inequalities are satisfied, i.e., both players have ``state-mismatching'' preferences, then with relabelling the actions, the analysis in the main text will go through.

\section{Comparative statics on preference misalignment induced by the principal}\label{app:cs:principal}

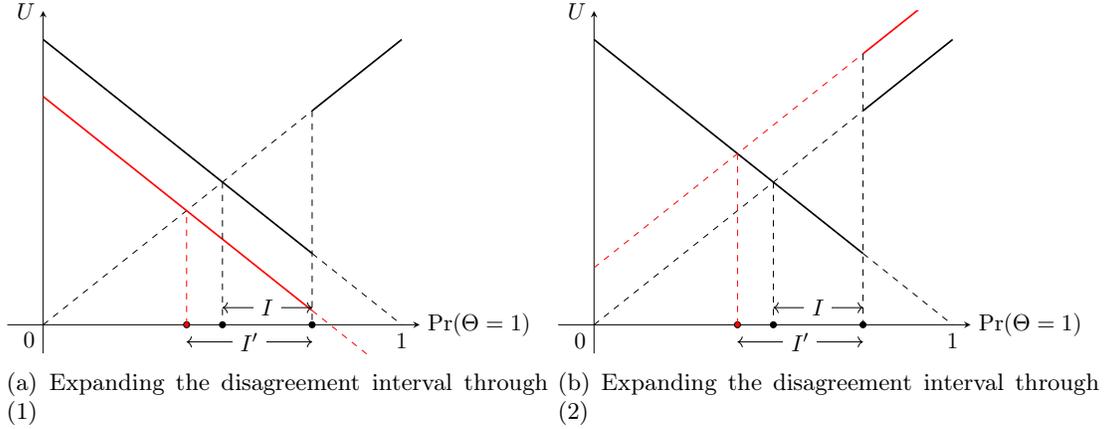
\begin{figure}[htb]
  \centering
  \subfigure[Expanding the disagreement interval through (1)]{  \begin{tikzpicture}[scale=0.8, transform shape]
    \pgfplotsset{compat=1.17}
    \begin{axis}[
      xlabel={$\Pr(\Theta=1)$},
      ylabel={$U$},
      xmin=-0.1, xmax=1.05,
      ymin=-0.1, ymax=1.1,
      axis lines=middle,
      samples=100,
      xtick=\empty,
      ytick=\empty,
      xlabel style={anchor=west},
      ylabel style={anchor=east},
      clip=true,
      ]
      \addplot[black, thick, domain=0.75:1] {x};
      \addplot[black, dashed, domain=0:0.75] {x};
      \addplot[black, thick, domain=0:0.75] {1 - x};
      \addplot[black, dashed, domain=0.75:1] {1 - x};
      \addplot[red, thick, domain=0:0.75] {0.8 - x};
      \addplot[red, dashed, domain=0.75:1] {0.8 - x};

      \coordinate (A) at (0.5, 0.5);
      \coordinate (B) at (0.75, 0.75);
      \coordinate (C) at (0.5, 0);
      \coordinate (D) at (0.75, 0);
      \coordinate (E) at (0.4, 0);

      \node[draw,circle,inner sep=1pt,fill=black] at (C) {};
      \node[draw,circle,inner sep=1pt,fill=black] at (D) {};
      \node[draw,circle,inner sep=1pt,fill=red] at (E) {};

      \node[below left] at (0, 0) {$0$};
      \node[below] at (1, 0) {1};
      \draw[dashed] (A) -- (0.5, 0);
      \draw[dashed] (B) -- (0.75, 0);

      \draw[dashed, red] (E) -- (0.4, 0.4);
      \draw[<->] (0.5,0.06) -- (0.75,0.06) node[midway, fill=white] {$I$};
      \draw[<->] (0.4,-0.06) -- (0.75,-0.06) node[midway, fill=white] {$I'$};
      
    \end{axis}
  \end{tikzpicture}}
  \subfigure[Expanding the disagreement interval through (2)]{  \begin{tikzpicture}[scale=0.8, transform shape]
    \pgfplotsset{compat=1.17}
    \begin{axis}[
      xlabel={$\Pr(\Theta=1)$},
      ylabel={$U$},
      xmin=-0.1, xmax=1.05,
      ymin=-0.1, ymax=1.1,
      axis lines=middle,
      samples=100,
      xtick=\empty,
      ytick=\empty,
      xlabel style={anchor=west},
      ylabel style={anchor=east},
      clip=true,
      ]
      \addplot[black, thick, domain=0.75:1] {x};
      \addplot[black, dashed, domain=0:0.75] {x};
      \addplot[black, thick, domain=0:0.75] {1 - x};
      \addplot[black, dashed, domain=0.75:1] {1 - x};
      \addplot[red, dashed, domain=0:0.75] {x+0.2};
      \addplot[red, thick, domain=0.75:1] {x+0.2};

      \coordinate (A) at (0.5, 0.5);
      \coordinate (B) at (0.75, 0.95);
      \coordinate (C) at (0.5, 0);
      \coordinate (D) at (0.75, 0);
      \coordinate (E) at (0.4, 0);

      \node[draw,circle,inner sep=1pt,fill=black] at (C) {};
      \node[draw,circle,inner sep=1pt,fill=black] at (D) {};
      \node[draw,circle,inner sep=1pt,fill=red] at (E) {};

      \node[below left] at (0, 0) {$0$};
      \node[below] at (1, 0) {1};
      \draw[dashed] (A) -- (0.5, 0);
      \draw[dashed] (B) -- (0.75, 0);

      \draw[dashed, red] (E) -- (0.4, 0.6);

      \draw[<->] (0.5,0.06) -- (0.75,0.06) node[midway, fill=white] {$I$};
      \draw[<->] (0.4,-0.06) -- (0.75,-0.06) node[midway, fill=white] {$I'$};
      
    \end{axis}
  \end{tikzpicture}}
  \caption{Principal-induced increase in preference misalignment}\label{fig:expanding}
\end{figure}

In this appendix, we consider the cases when the disagreement interval expands not because of a change in the agent's preferences but that of the principals. 
There are two ways this can happen, as shown in Figure \ref{fig:expanding}: (1) decrease the principal payoff for the agent's preferred action, and (2) increase the principal's payoff for the agent's less preferred action. 

\begin{proposition}
For any principal's interim posterior and any agent's signal, both the principal's payoff of delegation and payoff at the optimal delegation decision weakly decreases when the disagreement interval expands through (1), and weakly increases when the disagreement interval expands through (2).
\end{proposition}

\begin{proof}
Suppose that the original disagreement interval is $I$ and that it expands to $I' \supset I$ under the new agent. Denote the principal's original delegation envelope $V_D$ and new delegation envelope $V'_D$. Denote the principal's original nondelegation envelope $V_N$ and new nondelegation envelope $V'_N$.

We observe that in case (1), both the principal's delegation and nondelegation envelope is weakly lower than before, i.e., $V'_D(x)\leq V_D(x)$ and $V'_N(x)\leq V_N(x)$ for all $x$. Since the principal's payoff of delegation is a linear combination of two points on the delegation envelope, it is also weakly lower than before. Similarly, the principal's payoff of nondelegation is a linear combination of two points on the nondelegation envelope, which is weakly lower than before. The principal's payoff at the optimal delegation decision, being the maximum of the two, is thus weakly lower than before. 

Case (2) is symmetric in that both the principal's delegation and nondelegation envelope is weakly higher than before, i.e., $V'_D(x)\geq V_D(x)$ and $V'_N(x)\geq V_N(x)$ for all $x$. By a symmetric argument, both the principal's payoff of delegation and payoff at the optimal delegation decision is weakly higher than before.
\end{proof}

\section{Derivation of the principal's ex-ante delegation payoff}\label{app:special_q}

This appendix derives the general formula for the intermediate section of the principal's ex-ante delegation payoff function. It also shows that the intermediate section will be flat if the principal's payoffs are symmetric. 

Since we are considering the posteriors and signals in the delegation stage, we can simplify the notations to omit the mentioning of signal $s_1$. We redefine:
\[\mu \coloneqq \mu_{s_1}, \quad \mu_0 \coloneqq \mu_{s_10}, \quad \mu_1 \coloneqq \mu_{s_11}\]
Being in the intermediate section means that each point is produced by averaging two points on the left and right part of the principal's delegation envelope. Let $g$ be the principal's delegation envelope.
\[g(\mu_0) = a+b\mu_0, \quad g(\mu_1)=c+d\mu_1\]
The slope of the line segment connecting the two points is:
\[\frac{ a+b\mu_0-c-d\mu_1}{\mu_0 - \mu_1}\]
We can write out the function of the connecting line segment $h$:
\[h(x) = \frac{ a+b\mu_0-c-d\mu_1}{\mu_0 - \mu_1}(x-\mu_1) + c+d\mu_1\]
Evaluate $h$ at the prior $\mu$:
\[h(\mu) = \frac{ a+b\mu_0-c-d\mu_1}{\mu_0 - \mu_1}(\mu-\mu_1) + c+d\mu_1\]
It's useful to write the prior as an weighted average of two posteriors:
\[\mu = w_1 \mu_0 + (1-w_1) \mu_1, \quad w_1 = q_0(1-\mu) + (1-q_1)\mu\]
Substitute in $\mu$:
\begin{align*}
  h(\mu) &= \frac{ a+b\mu_0-c-d\mu_1}{\mu_0 - \mu_1}(w_1 \mu_0 + (1-w_1) \mu_1-\mu_1) + c+d\mu_1 \\
         &= \frac{ a+b\mu_0-c-d\mu_1}{\mu_0 - \mu_1}(\mu_0 -\mu_1)w_1 + c+d\mu_1 \\
         &= (a+b\mu_0-c-d\mu_1)w_1 + c+d\mu_1 \\
         &= c + (a-c + b\mu_0)w_1 + d\mu_1(1-w_1) \\
         &= c + (a-c + b\mu_0)w_1 + d(\mu - w_1\mu_0) \\
         &= c + (a-c + (b-d)\mu_0)w_1 + d\mu 
\end{align*}
Substitute in $\mu_0$ and $w_1$ and simplify:
\begin{align*}
  h(\mu) &= c+ (a-c)q_0 + (a-c)(1-q_0-q_1)\mu + (b-bq_1+dq_1)\mu
\end{align*}
The slope of $h(\mu)$ with respect to $\mu$ is thus:
\[ s = (a-c)(1-q_0-q_1) + (b-bq_1+dq_1) \]
We can also express them in terms of model primitives by substituting $a=r_{00}$, $b = r_{10}-r_{00}$, $c = r_{01}$, and $d=r_{11}-r_{01}$:
\[s = r_{10}-r_{01} +  (r_{01}-r_{00})q_0 + (r_{11}-r_{10})q_1 \]

When the principal's payoff is symmetric ($r_{01} = r_{10}$ and $r_{11} = r_{00}$) as well as when the public signal is symmetric ($q_0 = q_1$), we have that 
\[s = 0, \quad h(\mu) = q. \]

\section{The optimal delegation-stage payoff when $\rho > \bar{\mu}_P$}\label{app:plot}

\begin{figure}[htb]
  \centering
  \subfigure{\begin{tikzpicture}[scale=0.8, transform shape]
  \begin{axis}[
    xlabel={$\mu_{s_1}$},
    ylabel={$H$},
    xmin=-0.1, xmax=1,
    ymin=-0.1, ymax=1,
    axis lines=middle,
    samples=100,
    xtick=\empty,
    ytick=\empty,
    xlabel style={anchor=west},
    ylabel style={anchor=east},
    clip=false,
    ]
    \addplot[black, thick, domain=0.7:1] {x};
    \addplot[black, thick, domain=0.55:0.7] {0.7};
    \addplot[black, thick, domain=0:0.5] {1 - x};
    \addplot[black, thick, domain=0.5:0.55] {x};

    \coordinate (A) at (0.5, 0.5);
    \coordinate (B) at (0.75, 0.75);
    \coordinate (C) at (0.5, 0);
    \coordinate (D) at (0.75, 0);
    

    \node[below left] at (0, 0) {$0$};
    \node[below] at (1, 0) {1};

  \end{axis}
\end{tikzpicture}}
  \caption{Principal's optimal delegation-stage payoff}\label{fig:H2}
\end{figure}
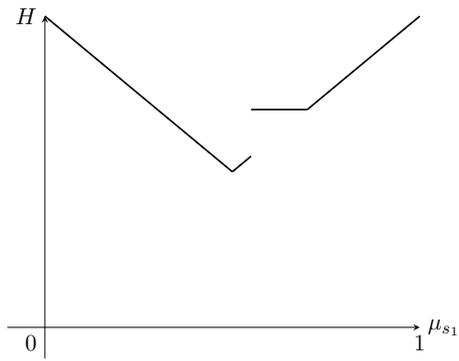

\end{document}